\documentclass[aps,pre,reprint,showpacs,superscriptaddress]{revtex4-1}
\usepackage{graphicx,amsmath,amssymb}
\usepackage[T1]{fontenc}

\begin{document}

\title{Infinite horizon billiards: \\ Transport at the border between Gauss and L\'evy universality classes}

\author{Lior Zarfaty}
\affiliation{Department of Physics, Institute of Nanotechnology and Advanced Materials, Bar-Ilan University, Ramat-Gan 52900, Israel}

\author{Alexander Peletskyi}
\affiliation{Institute of Physics, University of Augsburg, Universit\"atsstrasse 1, D-86135 Augsburg, Germany}

\author{Eli Barkai}
\affiliation{Department of Physics, Institute of Nanotechnology and Advanced Materials, Bar-Ilan University, Ramat-Gan 52900, Israel}

\author{Sergey Denisov}
\affiliation{Department of Computer Science, Oslo Metropolitan University, N-0130 Oslo, Norway}

\begin{abstract}

We consider transport in two billiard models, the infinite horizon Lorentz gas and the stadium channel, presenting analytical results for the spreading packet of particles. We first obtain the cumulative distribution function of traveling times between collisions, which exhibits non-analytical behavior. Using a renewal assumption and the L\'evy walk model, we obtain the particles' probability density. For the Lorentz gas, it shows a distinguished difference when compared with the known Gaussian propagator, as the latter is valid only for extremely long times. In particular, we show plumes of particles spreading along the infinite corridors, creating power-law tails of the density. We demonstrate the slow convergence rate via summation of independent identically distributed random variables on the border between L\'evy and Gauss laws. The renewal assumption works well for the Lorentz gas with intermediately sized scattering centers, but fails for the stadium channel due to strong temporal correlations. Our analytical results are supported with numerical samplings.

\end{abstract}

\maketitle

\section{Introduction}

The infinite horizon Lorentz gas \cite{Lorentz} is a paradigmatic model of deterministic classical transport, thoroughly studied by physicists \cite{Bouchaud,Zaslavsky,Cristadoro1,Cristadoro2,Pedro,Klages,Carlos} and mathematicians \cite{Bunimovich1,Boldrighini,Bunimovich2,Bleher,Dahlqvist,Szasz,Zaharescu,Chernov1,Chernov2,Chernov3,Marklof1,Dettmann1,Dettmann2,Marklof2}. It consists of an infinite periodic lattice of convex obstacles, and point-like particles which undergo elastic collisions with them. The most common configuration of the Lorentz gas model is composed of circular scatterers arranged into an infinite square lattice of unit spacing. Several important properties of this model, such as ergodicity \cite{Szasz} and algebraic decay of the velocity correlations in time \cite{Bunimovich2}, have been rigorously proven. Importantly, Bleher \cite{Bleher} showed that a particle’s position vector $\boldsymbol{r}(t)$ has a limiting Gaussian distribution when normalized correctly. More accurately, the quantity $\lim_{t\to\infty}[\boldsymbol{r}(t)-\boldsymbol{r}(0)]/\sqrt{t\ln(t)}$ is a two-dimensional Gaussian variable with zero mean and a covariance matrix which depends on the arrangement of scatterers. However, this asymptotic form is valid only when $\ln[\ln(N)]/\ln(N)=\epsilon\ll1$, where $N$ is the number of collisions. To satisfy this condition, $N$ has to be extremely large (e.g., $\epsilon=0.01$ requires that $N\approx10^{281}$). Bleher's time is too large to be physically relevant \cite{Zaslavsky}, while microscopic inter-collision times cannot describe transport processes happening on much larger time scales. As such, key features of the Lorentz gas model can only be seen for intermediate times that are most relevant for transport regimes. Unfortunately, currently there are no analytical results for these \textit{mesoscopic} times.

\begin{figure}
	\includegraphics[width=1.0\columnwidth]{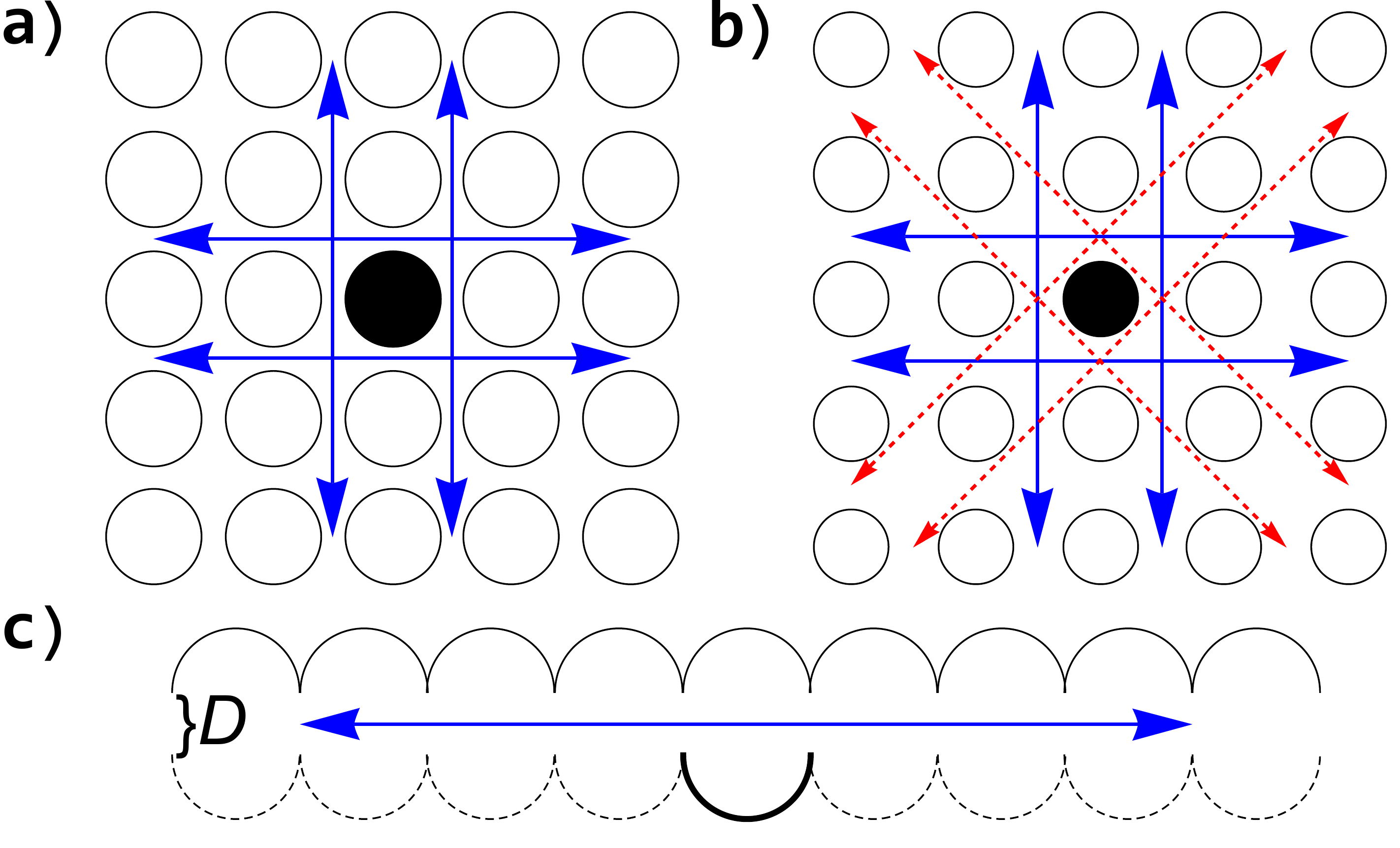}
	\caption{The considered models, the infinite horizon Lorentz gas with two (a) and four (b) open horizons, and the stadium channel model (c). For the Lorentz gas we take the lattice constant to be $1$, while for the stadium channel we use $1$ for the radius of a semicircle wall. The parameter which controls the qualitative behavior for the Lorentz gas is the scatterers' radius $R$. When $1/\sqrt{8} \le R < 1/2$, there are two directions of infinite corridors. A particle that collided with the black circle can fly along the blue solid arrows, reaching scatterers along them. Decreasing the radius such that $1/\sqrt{20} \le R < 1/\sqrt{8}$ creates additional two directions of infinite corridors. A particle that collided with the black circle can now fly along the red dashed arrows as well (see Fig.~\ref{LorentzSymmetry} for additional details). In the stadium channel model the controlling parameter is the walls' separation, $D>0$. A particle which was scattered from the thick bottom semicircle can reach any stadium of the top row, but can also hit the origin semicircle again (the dashed stadiums are unreachable in this case). All of the numerical simulations were performed using unit speed, namely $V=1$.}
\label{SetupSketches}
\end{figure}

\begin{figure*}
	\includegraphics[width=1.0\textwidth]{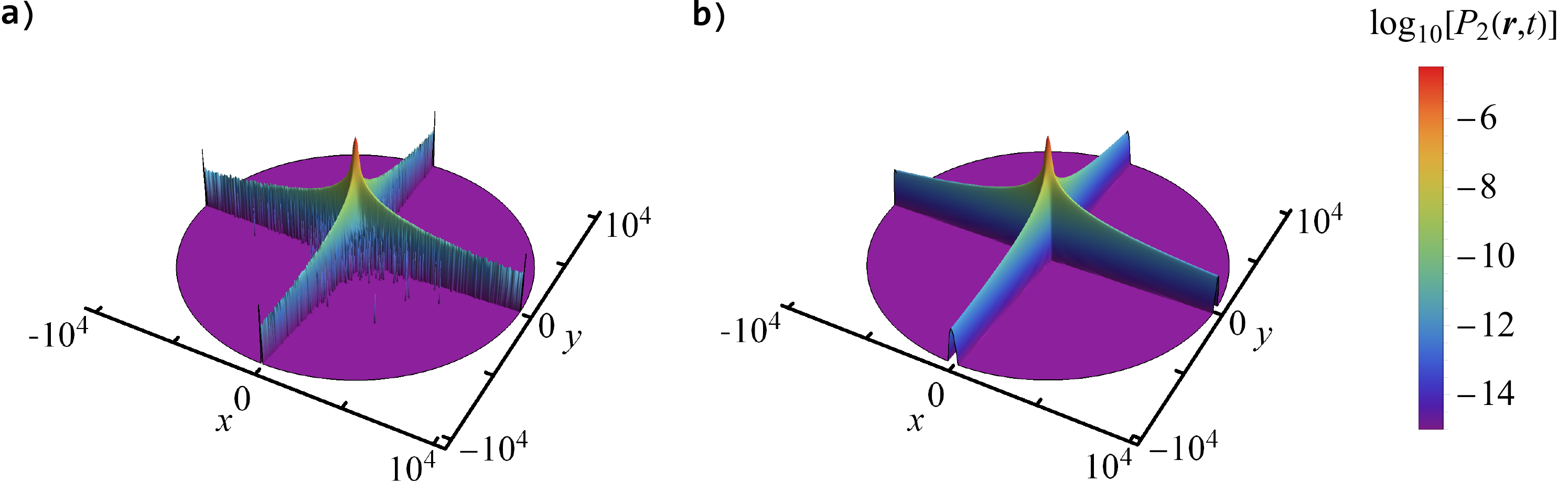}
	\caption{The position's probability density function for the Lorentz gas with two open horizons produces a cross-like geometry. Here, the lattice constant and speed are $1$, the scatterers' radius is $R=0.4$, and the time duration is $t=10^4$. The Lorentz gas simulation (a) has approximately $10^9$ sampled trajectories. Our theory Eq.~(\ref{equation: position distribution for the gas}) with $q=0$ reproduces the simulation well (b).}
	\label{LorentzCross3D}
\end{figure*}

\begin{figure*}
	\includegraphics[width=1.0\textwidth]{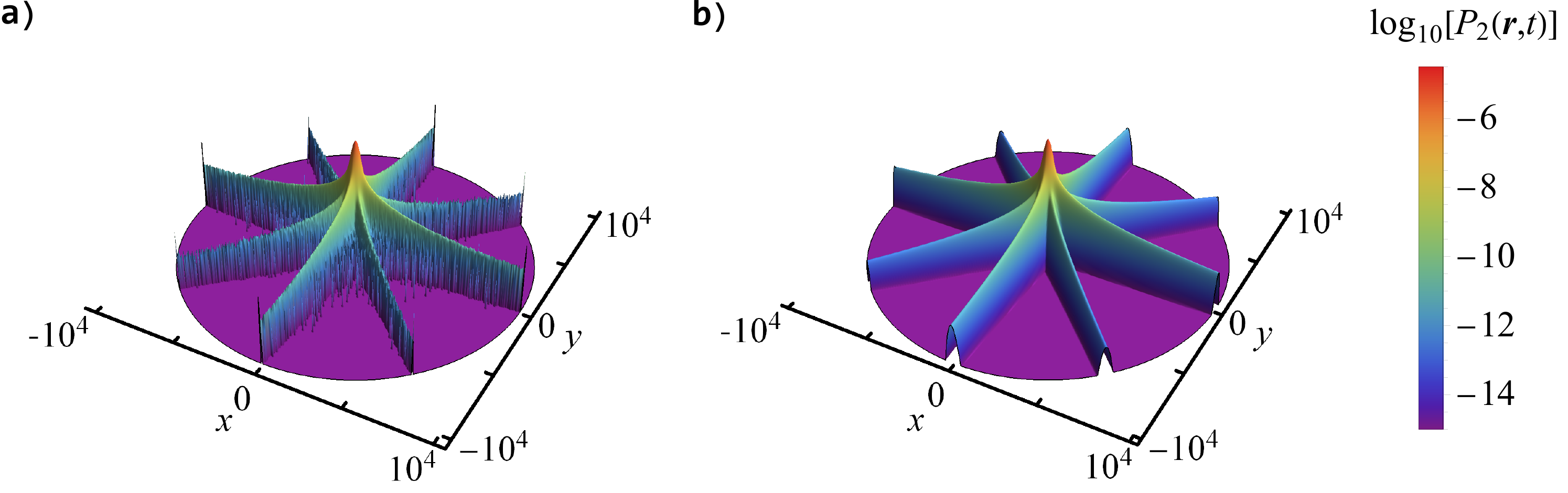}
	\caption{The position's probability density function for the Lorentz gas with four open horizons produces a British flag-like shape. Here, the lattice constant and speed are $1$, the scatterers' radius is $R=0.3$, and the time duration is $t=10^4$. The Lorentz gas simulation (a) is reproduced without any fitting by our theory (b), which is given by Eqs.~(\ref{equation: position distribution for the gas}) and (\ref{equation: positive q definition}).}
	\label{LorentzFlag3D}
\end{figure*}

\begin{figure}
	\includegraphics[width=1.0\columnwidth]{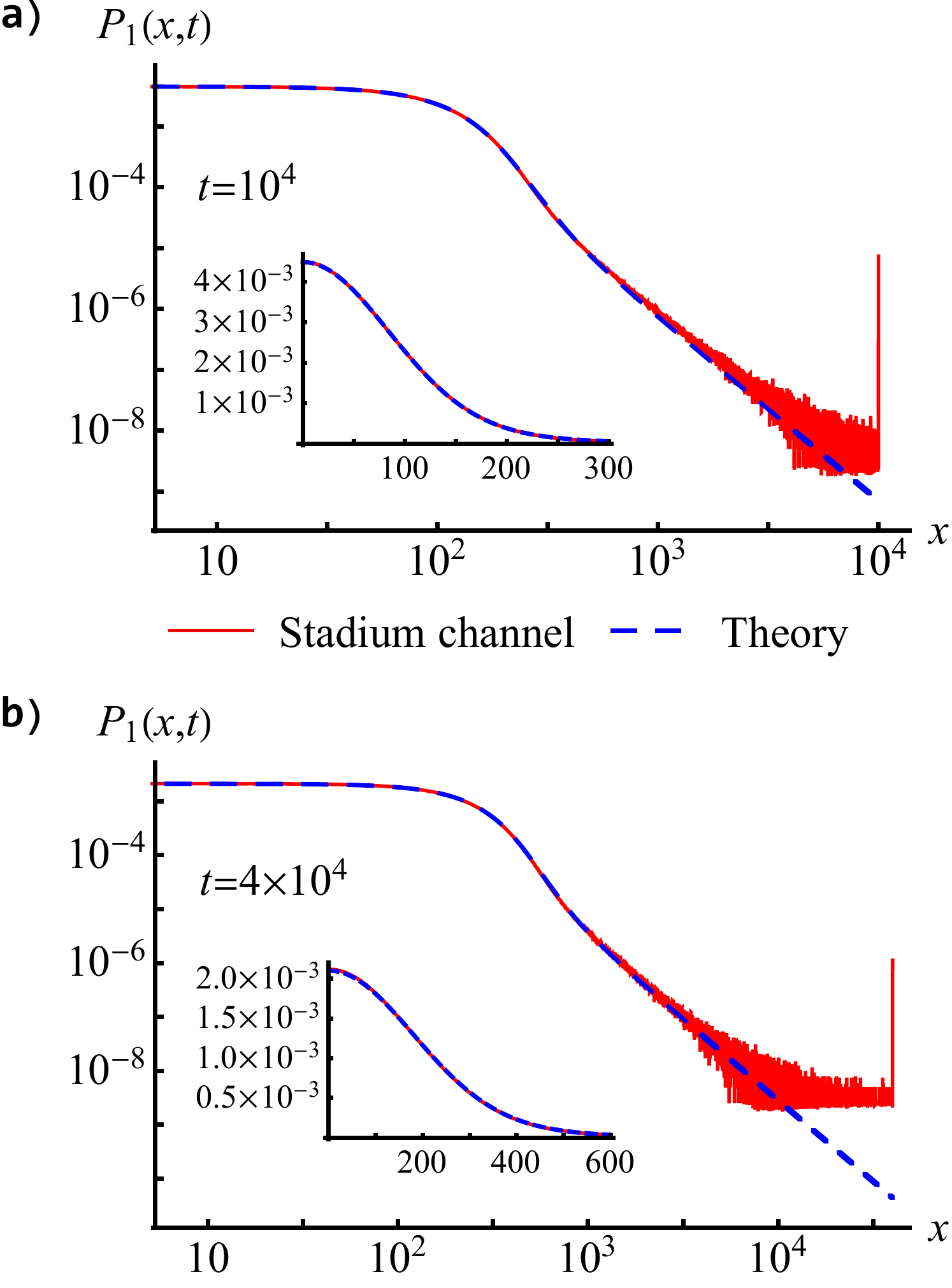}
	\caption{The position's probability density function of the stadium channel model for $t=10^4$ (a) and $t=4\times 10^4$ (b), with $D=1$. Solid red is the stadium channel numerical simulations, dashed blue is Eq.~(\ref{equation: position distribution for the pipe}). Using a two parameter fitting procedure and the simulation of duration $t=10^4$, we find that $C_{\psi}^2\langle\tau\rangle\approx0.3776$ and $\langle\tau\rangle/\tau_0^2\approx7.0607$. These values were used when calculating $P_1(x,t)$ for $t=4\times 10^4$. The curves match well, which means that Eq.~(\ref{equation: position distribution for the pipe}) can indeed describe the stadium channel model, given an ``effective" waiting times distribution. The stadium channel simulations have around $2\times 10^8$ sampled trajectories (the smallest values in the histograms represent cells with a single event).}
	\label{PipePositionPDF}
\end{figure}

In billiard models, an important characteristic of the transport is a presence (or an absence) of infinite horizons, corridors along which collision-less ballistic trajectories propagate, see Fig.~\ref{SetupSketches}. The packet of spreading particles in an \textit{infinite horizon} billiard models exhibits two main features: the center part of the packet is approximately Gaussian, and the far tails are described by plumes of particles spreading along the infinite corridors. For a specific configuration of the Lorentz gas, we found in our previous Rapid Communication \cite{PRE2018} that the scatterers' geometry is embedded in the cross-like shape of the spreading packet, see Figs.~\ref{SetupSketches} (a) and \ref{LorentzCross3D}. Here we wish to extend our theory to other configurations, showing its generality. For this aim, we consider the Lorentz gas with corridors forming a British flag-like structure for the packet of spreading particles, seen in Figs.~\ref{SetupSketches} (b) and \ref{LorentzFlag3D}, together with a quasi one-dimensional transport in a chain of concatenated Bunimovich billiard stadiums \cite{Bunimovich1,Zaslavsky}, seen in Figs.~\ref{SetupSketches} (c) and \ref{PipePositionPDF}. While the models are distinctive, along the corridors the far tails of the density decay spatially with a universal power-law, a feature well described by the L\'evy walk model \cite{Levy1,Levy2,Levy3,Levy4}.

Our analysis is composed of two main ingredients, one of them is obtaining the aforementioned billiard systems' full distribution of inter-collision times. For the Lorentz gas, results in the limits of $R \to 1$ and $\tau\to\infty$ (where $R$ is a scatterer's radius and $\tau$ is an inter-collision time) were found by Bouchaud and Le Doussal \cite{Bouchaud}. More recently, an asymptotic form in the limit of $R \to 0$ was established \cite{Dahlqvist,Zaharescu,Marklof1}. However, this important aspect in the characterization of transport is hardly discussed in the literature when finite sized scatterers are considered. This distribution's probability density function (PDF) exhibits a far tail obeying
\begin{equation}
\label{equation: fat tail}
	\lim_{\tau\to\infty}\tau^3\psi(\tau)=\tau_0^2 ,
\end{equation}
which is valid for both limits of large and small scatterers. Therefore, we use a L\'evy walk model with exponent $-3$ for the other ingredient, which is calculating the density of spreading particles. A principal issue here is that in a L\'evy walk approximation one uses a renewal assumption, i.e. one neglects correlations between consecutive collisions. It was shown that for the cross-like configuration of the Lorentz gas and in the limit of large scattering centers, this condition is nullified as an effective trapping mechanism emerges \cite{Cristadoro1}. In contrast, when the scattering centers are not too large, the L\'evy walk with the obtained cumulative distribution function (CDF) of the waiting times works perfectly, as demonstrated below. Deviations from the renewal theory do exist for the stadium channel model where correlations are strong, as we show below. However, this does not imply that the L\'evy walk model is not predictive here as well, and in fact we find the opposite. Another primary issue is caused since these two systems are operating on the border between Gauss and L\'evy central limit theorems, due to the exponent $-3$ which causes a logarithmic divergence of $\psi(\tau)$'s second moment. Thus, an ultra-slow convergence rate problem arises that can be understood via a toy model: summation of independent identically distributed (IID) random variables (RVs), with a common symmetric PDF that decays algebraically with a power of $-3$ for large argument. Already here we encounter the same type of convergence problem discussed in the first paragraph, which is neutralized using what we call Lambert scaling approach. This is a crucial step for these systems, as it allows us to compare finite time simulations with our theory for duration regimes where previous results do not hold.

The rest of this paper is organized as follows: in section II we provide an example of the Lambert scaling for sums of IID RVs. In section III we derive the main formula of the spatial PDFs by using the L\'evy walk model. In section IV we obtain the CDFs of inter-collision times for the Lorentz gas and stadium channel. We discuss our results in section V.

\section{A simplified case of Lambert scaling}
\label{section: simplified case of lambert}

\begin{figure}
	\includegraphics[width=1.0\columnwidth]{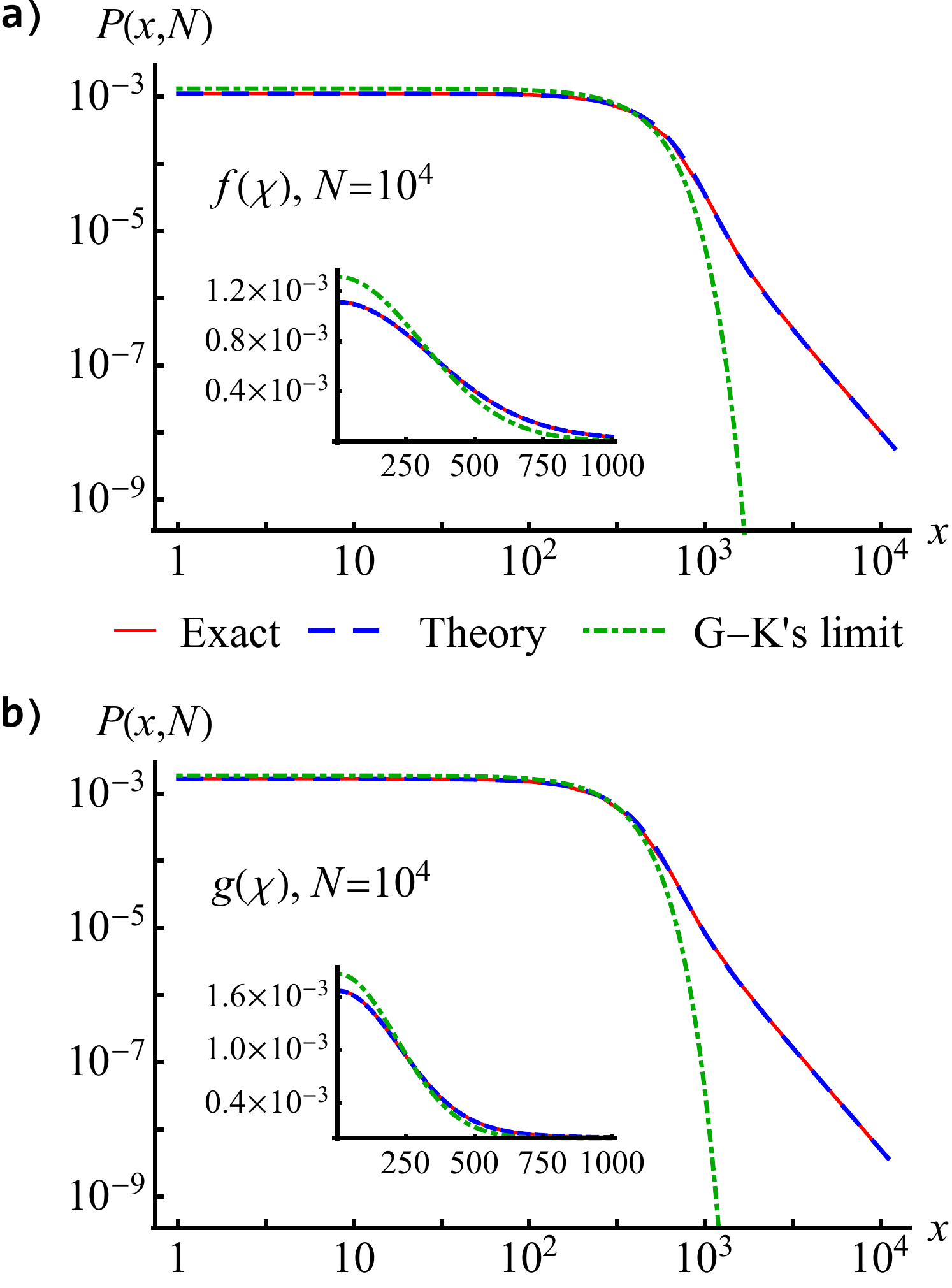}
	\caption{The probability density function (PDF) of the sum Eq.~(\ref{equation: iid sum}) with $N=10^4$, and its corresponding Lambert scaling approximation Eq.~(\ref{equation: pdf for simple case}). The common PDFs used here are $f(\chi)=1/|\chi|^3$ when $|\chi|\ge1$, and $0$ otherwise (a), and $g(\chi)=(1+\chi^2)^{-3/2}/2$ (b). The exact values (solid) are obtained from an inverse Fourier transform of $\tilde{f}^N(k)$, dashed lines are the theory Eq.~(\ref{equation: pdf for simple case}), and dot-dashed are the limit distributions of Gnedenko-Kolmogorov (G-K) \cite{Limit}.}
	\label{PDFofIID}
\end{figure}

We now consider the problem of summation of IID RVs, drawn from a power-law distribution. We work at the border between Gauss and L\'evy central limit theorems, which is clearly related to the exponent $-3$ in Eq.~(\ref{equation: fat tail}). Some aspects of this by far simpler approach are important for the discussed billiard models. In particular, at this transition we find a critical slowing down in the sense that convergence to the Gaussian limit theorem is ultra-slow \cite{Greengard}, a problem which is resolved by Lambert scaling. Consider a sum of $N\gg 1$ IID RVs
\begin{equation}
\label{equation: iid sum}
	x=\sum_{n=1}^N \chi_n ,
\end{equation}
where the summands are drawn from a common symmetric PDF which obeys $f(\chi\to\infty)\simeq\chi_0^2/\chi^3$. We define the scaled sum as $\bar{x}=x/\sqrt{\chi_0^2 N \Omega(N)/2}$ with $N\Omega(N)$ being a scaling parameter, soon to be determined. We use the characteristic function
\begin{equation}
\label{equation: characteristic function}
	\langle \exp( i \bar{k} \bar{x})\rangle = \exp\left\{ N \ln\left[ \tilde{f} \left( \frac{\bar{k}}{\sqrt{\chi_0^2 N\Omega(N)/2}}\right) \right] \right\},
\end{equation}
where $\tilde{f}(k)$ is the Fourier transform of $f(\chi)$. Assuming that $\Omega(N)$ monotonically increases with $N$, the small $k$ behavior of $\tilde{f}(k)$ is considered
\begin{equation}
\label{equation: small k behavior}
	\tilde{f}(k)\simeq 1+\frac{1}{2}\left(\chi_0k\right)^2 \ln\left[\left(C_f\chi_0k\right)^2\right] ,
\end{equation}
which is derived in appendix \ref{appendix: derivation small k}. The first term is the normalization, while the second is related to the power-law tail of $f(\chi)$, with $C_f$ being
\begin{align}
\label{equation: Cf definition}
	C_f = \exp\left\{\vphantom{\int_{\chi_0}^{\infty} \text{d}\chi \left[ f(\chi)\left(\frac{\chi}{\chi_0}\right)^2 - \frac{1}{\chi}\right]} \gamma \right. & - \frac{3}{2} - \int_{0}^{\chi_0} \text{d}\chi \, f(\chi) \left(\frac{\chi}{\chi_0}\right)^2 \\
	&- \left. \int_{\chi_0}^{\infty} \text{d}\chi \left[ f(\chi)\left(\frac{\chi}{\chi_0}\right)^2 - \frac{1}{\chi}\right] \right\} \nonumber ,
\end{align}
where $\gamma \approx 0.5772$ is Euler's constant. Inserting Eq.~(\ref{equation: small k behavior}) into Eq.~(\ref{equation: characteristic function}) and expanding, we get
\begin{equation}
\label{equation: characteristic function expansion 1}
	\left< \exp( i \bar{k} \bar{x} ) \right> \simeq \exp\left\{ \frac{\bar{k}^2}{\Omega(N)} \ln\left[ \frac{2C_f^2\bar{k}^2}{N\Omega(N)} \right] \right\} .
\end{equation}
We now determine the slowly increasing function $\Omega(N)$ with the choice
\begin{equation}
\label{equation: lambert scaling choice}
	\ln\left[\frac{N\Omega(N)}{2C_f^2}\right] = \Omega(N) ,
\end{equation}
which yields
\begin{equation}
\label{equation: omega definition for N}
	\Omega(N) = \left|\text{W}_{-1} \left( - \frac{2C_f^2}{N} \right)\right|. 
\end{equation}
Here $\text{W}_{-1}(\eta)$ is the secondary branch of the Lambert W-function \cite{Functions}, defined for $\eta\in[-1/e,0)$ by the identity $\text{W}_{-1}(\eta) = \ln[ \eta /\text{W}_{-1}(\eta)]$. The Lambert function has the following expansion as $\eta\rightarrow 0^-$
\begin{equation}
\label{equation: lambert behavior}
	\left|\text{W}_{-1} (\eta)\right| = L_1 + L_2 + \frac{L_2}{L_1} + \mathcal{O}\left(\frac{L_2^2}{L_1^2}\right) ,
\end{equation}
where $L_1=\ln(1/|\eta|)$ and $L_2=\ln[\ln(1/|\eta|)]$. Equations~(\ref{equation: omega definition for N}) and (\ref{equation: lambert behavior}) yield $N\Omega(N)\simeq N\ln(N)$ when $N\to\infty$, reproducing the well-known Gnedenko-Kolmogorov scaling $\sqrt{N\ln(N)}$ \cite{Limit}. However, for this to be of relevance, one must demand $\ln(N)\gg\ln[\ln(N)]$, which makes the convergence to this mathematical limit ultra-slow. Our Lambert scaling approach unites all of the $N$-dependent logarithmic terms into a single function, thus resolving this problem. We therefore expand Eq.~(\ref{equation: characteristic function expansion 1}) for large $\Omega(N)$, keeping terms up to sub-leading order during the calculation
\begin{equation}
\label{equation: characteristic function expansion 2}
	\left< \exp( i \bar{k} \bar{x} ) \right> \simeq e^{-\bar{k}^2} \left[ 1 + \frac{\bar{k}^2}{\Omega(N)} \ln\left( \bar{k}^2\right) \right] .
\end{equation}
Inverting Eq.~(\ref{equation: characteristic function expansion 2}) back to position space, we reach our first result, the PDF of $x$
\begin{align}
\label{equation: pdf for simple case}
	&P(x,N) \simeq \frac{1}{\sqrt{\pi\xi^2(N)}} \exp\left[ - \frac{x^2}{\xi^2(N)} \right] \times \nonumber \\
	&\left\{ 1 + \frac{1}{\Omega(N)} \left\{ \left[ 2 - \gamma - \ln(4) \vphantom{\frac{1}{2}}\right] \left[ \frac{1}{2} - \frac{x^2}{\xi^2(N)} \right] \right. \right. \nonumber \\
	&\left.\left. -\frac{1}{2} \text{M}^{(1,0,0)}\left[ - 1 ; \frac{1}{2} ; \frac{x^2}{\xi^2(N)} \right] \right\} \right\} ,
\end{align}
with $\xi(N)=\sqrt{2\chi_0^2N\Omega(N)}$. Here $\text{M}(\cdots)$ is Kummer's confluent hypergeometric function \cite{Functions}, and the superscript over $\text{M}$ denotes its derivative with respect to its first argument. Figure~\ref{PDFofIID} shows a good match between Eq.~(\ref{equation: pdf for simple case}) and the sum Eq.~(\ref{equation: iid sum})'s PDF for two different common density functions, $f(\chi)$ and $g(\chi)$ which are defined in the caption, with $N=10^4$. The Kummer function behave asymptotically as $\text{M}^{(1,0,0)}(-1,1/2,\eta^2) \simeq -\sqrt{\pi}\exp(\eta^2)/\eta^3$, suggesting that $P(x,N) \simeq N\chi_0^2/x^3$. Thus, the PDF of the sum $x$ reproduces the same power-law tails as of the original common PDF $f(\chi)$. As the distribution of traveling times between collisions Eq.~(\ref{equation: fat tail}) exhibit the same heavy-tail exponent of $-3$, our next step is implementing Lambert scaling to the L\'evy walk model.

\section{Lambert scaling of the L\'evy walk model}
\label{section: labert scaling for levy walk}

The $d$-dimensional L\'evy walk model \cite{Levy2,Levy4} is defined as follows: a random walker is placed at $\boldsymbol{r}(0)$ on time $t=0$. Its movement consists of segments of ballistic motion with constant velocity, separated by collision-like events which induce a change in the velocity's magnitude and/or direction. The process lasts for a fixed duration, which is the measurement time $t$. The model employs two PDFs in order to determine the displacement during each of the ballistic motion epochs. The velocity of each segment is drawn from a PDF $F_d(\boldsymbol{v})$, whose moments are all finite, and is further assumed to be symmetric with respect to each of the components $v_j$ where $1\le j\le d$ (such that its odd moments vanish). The time duration of each ballistic section is drawn from a PDF $\psi(\tau)$. The movement continues until the allotted measurement time is met, thus the number of collisions $N$ in $[0,t]$ is random. This yields the total displacement as
\begin{equation}
	\boldsymbol{r}(t) - \boldsymbol{r}(0) = \sum_{n=1}^{N} \boldsymbol{v}_{n-1}\tau_n + \boldsymbol{v}_{N}\tau_b ,
\end{equation}
where $\tau_n$ is the traveling time of the $n$th walking epoch, $\boldsymbol{v}_n$ is the velocity after the $n$th collision, and the initial conditions $\boldsymbol{r}(0)$ and $\boldsymbol{v}_0$ are randomly chosen. The traveling times and velocities $\{\tau_n,\boldsymbol{v}_n\}$ (with $1\le n\le N$) are IID RVs, with the last movement duration being $\tau_b=t-\sum_{n=1}^{N}\tau_n$. Notice that the measurement time divided by the mean time between collisions, $t/\langle\tau\rangle$, and the lengths of inter-collision travel, $\{\boldsymbol{v}_{n-1} \tau_n\}$, roughly correspond to $N$ and $\{\chi_n\}$ from the previous section, respectively. We denote the particle's speed with $V$, which is kept unchanged in the billiards systems due to the collisions' elasticity (in the numerical simulations $V=1$). Let us denote the probability to find the walker at position $\boldsymbol{r}$ on time $t$ as $P_d(\boldsymbol{r},t)$, and let $\Pi_d(\boldsymbol{k},u)$ be its Fourier and Laplace transform
\begin{equation}
	\Pi_d\left(\boldsymbol{k}, u\right) = \int\text{d}^dr \int_0^{\infty}\text{d}t P_d\left(\boldsymbol{r},t\right) e^{-ut+i\boldsymbol{k}\cdot\boldsymbol{x}} .
\end{equation}
An exact expression of $\Pi_d(\boldsymbol{k},u)$ is given by the Montroll-Weiss equation \cite{MW}
\begin{equation}
\label{equation: montroll weiss}
	\Pi_d\left(\boldsymbol{k}, u\right) = \left< \frac{1 - \hat{\psi}\left(u-i\boldsymbol{k} \cdot\boldsymbol{v}\right)}{u-i\boldsymbol{k}\cdot\boldsymbol{v}} \right> \frac{1}{1 - \left< \hat{\psi} \left( u - i \boldsymbol{k} \cdot \boldsymbol{v} \right) \right> } ,
\end{equation}
where $\hat{\psi}(u)$ is the Laplace transform of $\psi(\tau)$, defined as
\begin{equation}
	\hat{\psi}(u) = \int_{0}^{\infty} \text{d}\tau \, \psi(\tau) e^{-u\tau} ,
\end{equation}
and
\begin{equation}
	\left<\cdots\right> = \int \text{d}^dv \cdots F_d(\boldsymbol{v}) ,
\end{equation}
with the above integral carried over all velocity space.

We now direct the reader's attention to two points before advancing with the calculation. First, notice that the assumption of finite moments for the velocity distribution $F_d(\boldsymbol{v})$ is crucial to our theory. For example, a diverging second moment yields L\'evy statistics in the bulk instead of a Gaussian, whereas for the Lorentz gas it is rigorously proven to be the latter case \cite{Bleher}. Second, in Ref.~\cite{Levy4} the authors discuss two different variations of L\'evy walk: the velocity model and the jump model. For the former, particles move constantly until the measurement time ends, their last traveling epoch being $\tau_b$. However, for the latter particles are missing this final segment of walk. Therefore, the two models differ only when considering approximation theories of the far tail (under the condition of finite mean time between collisions). As our theory is intended to approximate the bulk of the PDF $P(\boldsymbol{r},t)$, these two aforementioned cases are indistinguishable. Technically speaking, this difference is manifested by a different numerator of Eq.~(\ref{equation: montroll weiss}), which does not change its approximated versions which appear below, Eqs.~(\ref{equation: fourier transform of the pipe distribution}) and (\ref{equation: fourier transform of the gas distribution}). Further, in the Lorentz gas a particle’s velocity is unity at any moment of travel, therefore the velocity model is more suitable if compared to the jump model.

To approximate Eq.~(\ref{equation: montroll weiss}) we use the asymptotic behavior of $\psi(\tau)$ Eq.~(\ref{equation: fat tail}), which implies that for small $u$
\begin{equation}
\label{equation: small u behavior}
	\hat{\psi}(u) \simeq 1 - \langle \tau \rangle u - \frac{1}{2} \left( \tau_0 u\right)^2 \ln\left( C_{\psi} \tau_0 u \right) ,
\end{equation}
which is derived in appendix \ref{appendix: derivation small u}. Here, the first term is the normalization, $\langle\tau\rangle$ is the mean time between collisions, and the last term is related to the power-law tail of $\psi(\tau)$, with $C_{\psi}$ being
\begin{align}
\label{equation: Cpsi definition}
	C_{\psi} = \exp\left\{\vphantom{\int_{\tau_0}^{\infty} \text{d}\tau \left[ \psi(\tau)\left(\frac{\tau}{\tau_0}\right)^2 - \frac{1}{\tau}\right]} \gamma \right. & - \frac{3}{2} - \int_{0}^{\tau_0} \text{d}\tau \, \psi(\tau) \left(\frac{\tau}{\tau_0}\right)^2 \\
	&- \left. \int_{\tau_0}^{\infty} \text{d}\tau \left[ \psi(\tau)\left(\frac{\tau}{\tau_0}\right)^2 - \frac{1}{\tau}\right] \right\} . \nonumber
\end{align}
We assume a scaling of $u \sim k^2 L(k)$, where $L(\cdots)$ is some logarithmic-like function and $k=|\boldsymbol{k}|$. This suggests that $u \ll k$ when $k \rightarrow 0$, and as such we can expand the Montroll-Weiss Eq.~(\ref{equation: montroll weiss}) in the small parameter $u/kv$ where $v=|\boldsymbol{v}|$, see appendix \ref{appendix: additional details for lambert levy}. For the one-dimensional stadium channel model, we apply the following distribution of velocities
\begin{equation}
\label{equation: velocity distibution for the pipe}
	F_1\left(v_x\right) = \frac{1}{2} \left[ \delta\left(v_x-V\right) + \delta\left(v_x+V\right) \vphantom{\frac{1}{2}} \right] ,
\end{equation}
where in our case $V=1$. We obtain in appendix \ref{appendix: additional details for lambert levy} the following Fourier expansion for the distribution
\begin{equation}
\label{equation: fourier transform of the pipe distribution}
	\tilde{P}_1(\kappa_x,t) \simeq \frac{2e^{-\kappa_x^2}}{\xi_1(t)} \left[ 1 + \frac{1}{\Omega_1(t)} \kappa_x^2 \ln\left( \kappa_x^2 \right) \right] ,
\end{equation}
with
\begin{align}
\label{equation: omega and xi definition for t}
	&\Omega_d(t) = \left| \text{W}_{-1}\left( - 4dC_{\psi}^2 \frac{\left<\tau\right>}{t} \right) \right| , \nonumber \\
	&\xi_d(t) = \sqrt{\tau_0^2\langle v^2 \rangle \frac{t\Omega_d(t)}{d\langle\tau\rangle}} ,
\end{align}
where here and for the following Lorentz gas $\langle v^2 \rangle = V^2$. The walker's position PDF in $d$ spatial dimensions is given by the inverse Fourier transform
\begin{equation}
\label{equation: inverse fourier transform}
	P_d(\boldsymbol{r},t) = \int \frac{\text{d}^d\kappa}{(2\pi)^d} \tilde{P}_d\left(\boldsymbol{\kappa},t\right) \cos\left[ \frac{2(\boldsymbol{\kappa} \cdot \boldsymbol{r})}{\xi_d(t)} \right] .
\end{equation}
Inserting Eq.~(\ref{equation: fourier transform of the pipe distribution}) into Eq.~(\ref{equation: inverse fourier transform}) results with
\begin{align}
\label{equation: position distribution for the pipe}
	P_1(x,t) &= \frac{1}{\sqrt{\pi\xi_1^2(t)}} \exp\left[-\frac{x^2}{\xi_1^2(t)}\right] \nonumber \\
	&\times \left\{ 1 + \frac{1}{\Omega_1(t)} \left\{ \left[2-\gamma-\ln(4) \vphantom{\frac{x^2}{\xi_1^2(t)}}\right]\left[ \frac{1}{2}- \frac{x^2}{\xi_1^2(t)}\right] \right.\right.\nonumber\\
	&-\left.\left.\frac{1}{2}\text{M}^{(1,0,0)}\left[ -1;\frac{1}{2}; \frac{x^2}{\xi_1^2(t)}\right]\right\}\right\} .
\end{align}
The subscript $1$ stands for one dimension, namely this result should hold for the stadium channel when we coarse grain over the channel's width. We see some similarities with the problem of summation of IID RVs. For example, Kummer's function appears in both problems as a correction to the leading term. There is however a major difference between the two cases: here $P(\boldsymbol{r},t)=0$ for $\boldsymbol{r}>t$, which is different from the problem of summation of IID RVs. This is clearly due to the particles' finite speed. For the two-dimensional Lorentz gas model with two/four infinite horizons we have
\begin{align}
\label{equation: velocity distibution for the gas}
	F_2\left(\boldsymbol{v}\right) &= \frac{1-q}{4} \left\{ \left[ \delta\left(v_x-V\right) + \delta\left(v_x+V\right) \vphantom{\frac{v_{\times}}{\sqrt{2}}} \right] \delta\left(v_y\right) \right. \nonumber \\
	&+ \left. \delta\left(v_x\right) \left[ \delta\left(v_y-V\right) + \delta\left(v_y+V\right) \vphantom{\frac{v_{\times}}{\sqrt{2}}} \right] \right\} \nonumber \\
	&+ \frac{q}{4} \left\{ \left[ \delta\left(v_x-\frac{V}{\sqrt{2}}\right) + \delta\left(v_x+\frac{V}{\sqrt{2}}\right) \right] \right. \nonumber \\
	&\times \left. \left[ \delta\left(v_y-\frac{V}{\sqrt{2}}\right) + \delta\left(v_y+\frac{V}{\sqrt{2}}\right) \right] \right\} ,
\end{align}
where $q\ge 0$ is a parameter to be determined later, which encodes the probability of a particle to be found in the far tail of the diagonal corridors (for two open horizons, a cross-shape, one has $q=0$). We get from Eq.~(\ref{equation: montroll weiss})
\begin{align}
\label{equation: fourier transform of the gas distribution}
	\tilde{P}_2(\boldsymbol{\kappa},t) &\simeq \frac{4e^{-\kappa_x^2-\kappa_y^2}}{\xi_2^2(t)} \nonumber \\
	&\times \left\{ 1 + \frac{1-q}{\Omega_2(t)} \left[ \kappa_x^2 \ln\left( \kappa_x^2 \right) + \kappa_y^2 \ln\left( \kappa_y^2 \right) \vphantom{\frac{1}{2}} \right] \right. \nonumber \\
	 &+ \frac{q}{2\Omega_2(t)} \left(\kappa_x+\kappa_y\right)^2 \ln\left[ \frac{1}{2}\left(\kappa_x+\kappa_y\right)^2 \right] \nonumber \\
	 &+ \left. \frac{q}{2\Omega_2(t)} \left(\kappa_x-\kappa_y\right)^2 \ln\left[ \frac{1}{2}\left(\kappa_x-\kappa_y\right)^2 \right] \right\} .
\end{align}
We use a $\pi/4$ rotation transformation $\kappa_{\pm}=(\kappa_x\pm\kappa_y)/\sqrt{2}$ to calculate the integrals over Eq.~(\ref{equation: fourier transform of the gas distribution})'s last two terms, and obtain
\begin{align}
\label{equation: position distribution for the gas}
	P_2(\boldsymbol{r},t) &\simeq \frac{1}{\pi\xi_2^2(t)} \exp\left[-\frac{x^2+y^2}{\xi_2^2(t)}\right] \nonumber \\
	&\times \left\{ 1 + \frac{1}{\Omega_2(t)} \left\{ \left[2-\gamma-\ln(4) \vphantom{\frac{x^2}{\xi_2^2(t)}}\right]\left[ 1- \frac{x^2+y^2}{\xi_2^2(t)}\right] \right.\right.\nonumber\\
	&- \frac{1-q}{2}\text{M}^{(1,0,0)}\left[ -1;\frac{1}{2}; \frac{x^2}{\xi_2^2(t)}\right] \nonumber \\
	&-\frac{1-q}{2}\text{M}^{(1,0,0)}\left[ -1;\frac{1}{2}; \frac{y^2}{\xi_2^2(t)}\right] \nonumber \\
	&- \frac{q}{2}\text{M}^{(1,0,0)}\left[ -1;\frac{1}{2}; \frac{(x+y)^2}{2\xi_2^2(t)}\right] \nonumber \\
	& \left.\left. - \frac{q}{2}\text{M}^{(1,0,0)}\left[ -1;\frac{1}{2}; \frac{(x-y)^2}{2\xi_2^2(t)}\right]\right\}\right\} ,
\end{align}
where the subscript $2$ stands for two dimensions. This solution is presented in Figs.~\ref{LorentzCross3D} and \ref{LorentzFlag3D}, with the relevant parameters, namely $\tau_0$, $\langle\tau\rangle$, $C_{\psi}$ and $q$, obtained from $\psi(\tau)$ [more precisely we extract them out of $\text{CDF}(\tau)$]. Thus, we continue with deriving exact expressions for the distribution of traveling times $\psi(\tau)$ for the Lorentz gas model with two/four open infinite corridors and for the stadium channel model. To refrain from cumbersome formulas, we omit some of the next section's derivations. Please refer to appendix \ref{appendix: integration boundaries calculation} for more details.

\begin{figure}
	\includegraphics[width=1.0\columnwidth]{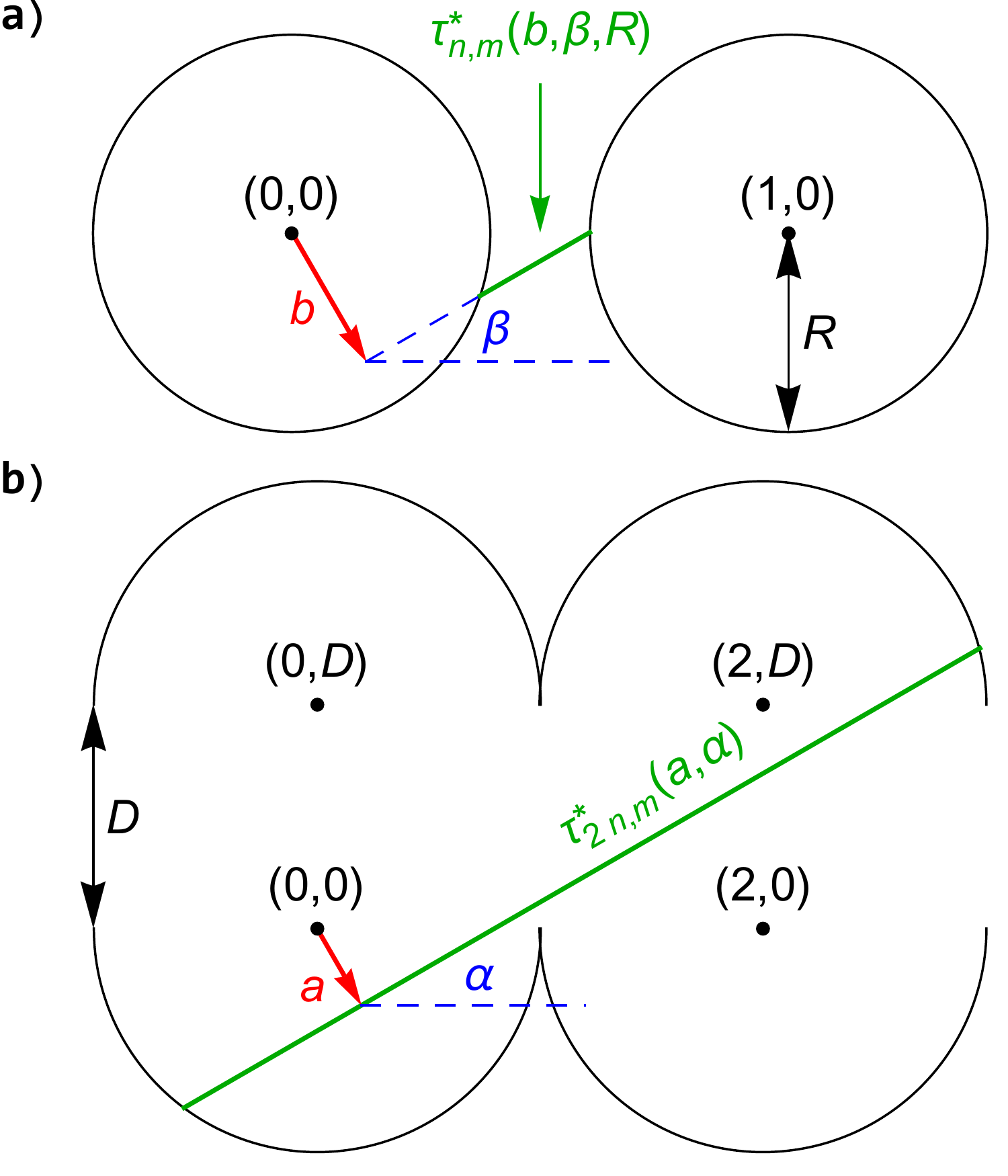}
	\caption{The parameters which determine the trajectory's length for the Lorentz gas model (a) and stadium channel model (b). For the Lorentz gas, the lattice coordinates of the next scatterer to be collided with are denoted as $(n,m)$, the recoil angle after the previous collision is $\beta$, and the impact parameter is $b$. In this sketch, we used $n=1$, $m=0$, $\beta=\pi/6$, $b=-0.3$ and $R=0.4$, which correspond to $\tau_{n,m}^*(b,\beta,R)\approx 0.255$ (the lattice constant is $1$). For the stadium channel, we denote as $(2n,m)$ the lattice site of the next stadium to be collided with, where $\alpha$ is the recoil angle after the previous collision and $a$ is an ``impact parameter". Here we used $n=1$, $m=D$, $\alpha=\pi/6$, $a=-0.4$ and $D=1$, which correspond to $\tau_{2n,m}^*(a,\alpha)\approx 4.11$ (the semicircles' radius is $1$).}
	\label{ParametersDefinitionsCDF}
\end{figure}

\begin{figure}
	\includegraphics[width=1.0\columnwidth]{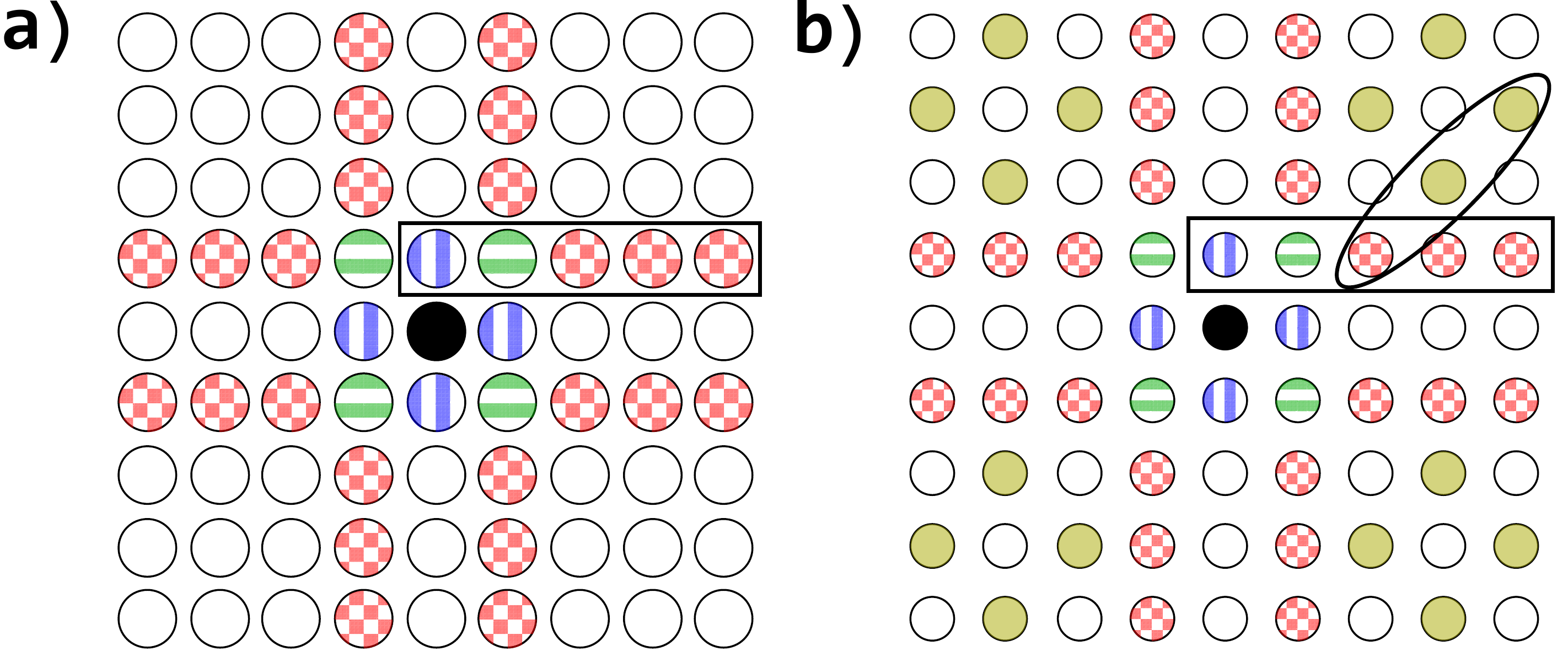}
	\caption{In the Lorentz gas, a tracer particle right after a collision with the origin circle (solid black) can hit only the colored scatteres. These are made of three distinct groups: four nearest neighbors (vertical blue), four next to nearest neighbors (horizontal green), and eight clusters of distant neighbors (chess red). The white circles are inaccessible for the particle. Given a lattice constant $1$, for a scatterers' radius of $R=0.4$ (a) one has two directions for possible infinite trajectories, directed with the lattice axes. For a scatterers' radius of $R=0.3$ (b) two diagonal infinite directions are added, and the particle can now reach the yellow scatterers (solid light gray in print). When calculating the cumulative distribution function of the traveling times between collisions, we sum the contribution of each scatterer to the trajectories' space, see section \ref{section: cdf calculation}. Exploiting the noticeable symmetry, we focus on the black rectangle-encircled area (a), and additionally the black ellipse-encircled area when $R$ is decreased (b).}
	\label{LorentzSymmetry}
\end{figure}

\section{The distribution of traveling times}
\label{section: cdf calculation}

\subsection{Lorentz gas model}

\begin{figure}
	\includegraphics[width=1.0\columnwidth]{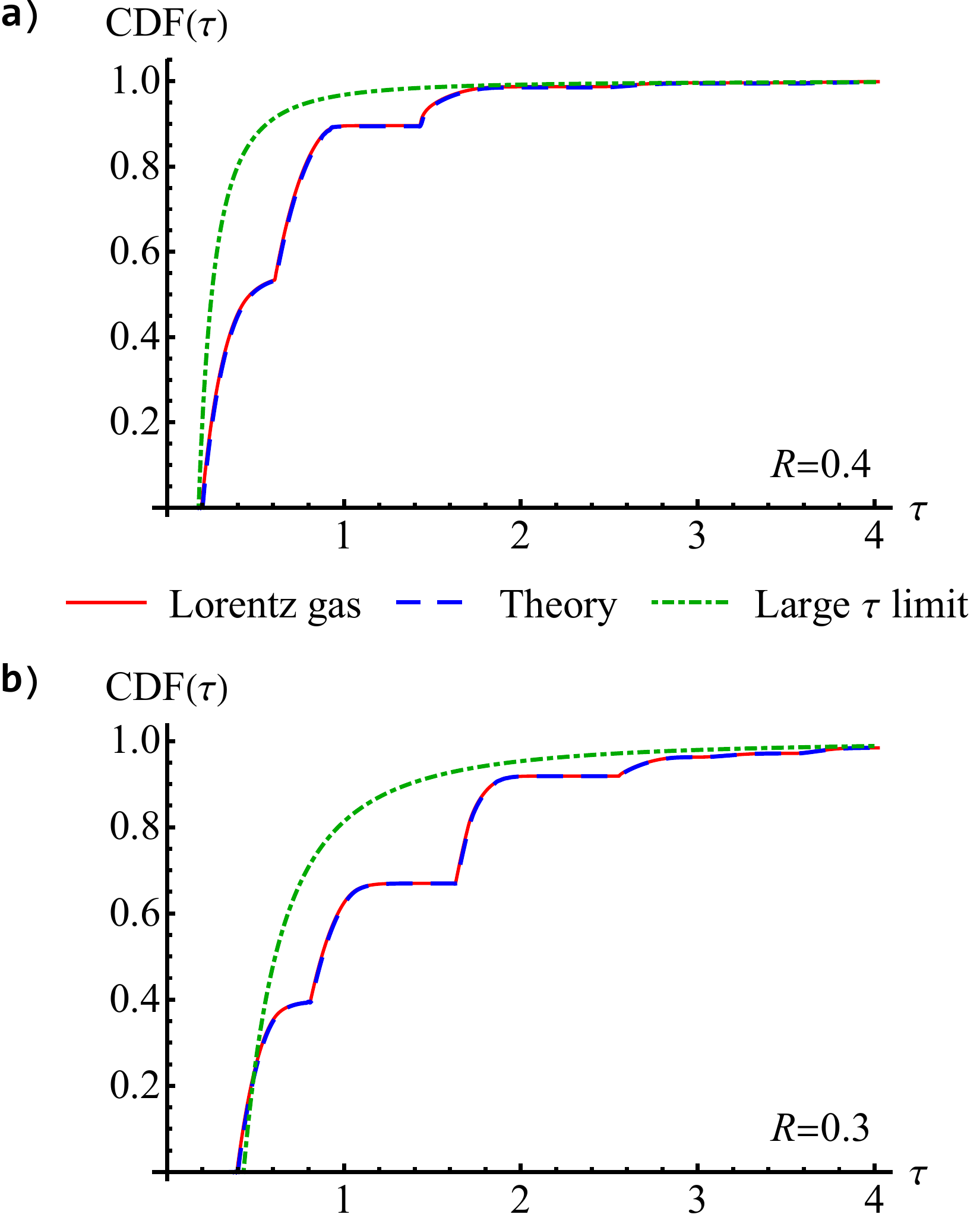}
	\caption{The cumulative distribution function of inter-collision times for the Lorentz gas model derived in section \ref{section: cdf calculation}, its respective result obtained from the numerical simulations, and the large $\tau$ limit given by $\text{CDF}(\tau)\simeq 1-\tau_0^2/2\tau^2$. The scatterers radius is $R=0.4$ (a) and $R=0.3$ (b), corresponding to two and four open infinite corridors, respectively (the lattice constant is $1$). The numerical histograms are made of a single long trajectory containing $10^6$ collisions.}
	\label{LorentzTauCDF}
\end{figure}

\begin{figure}
	\includegraphics[width=1.0\columnwidth]{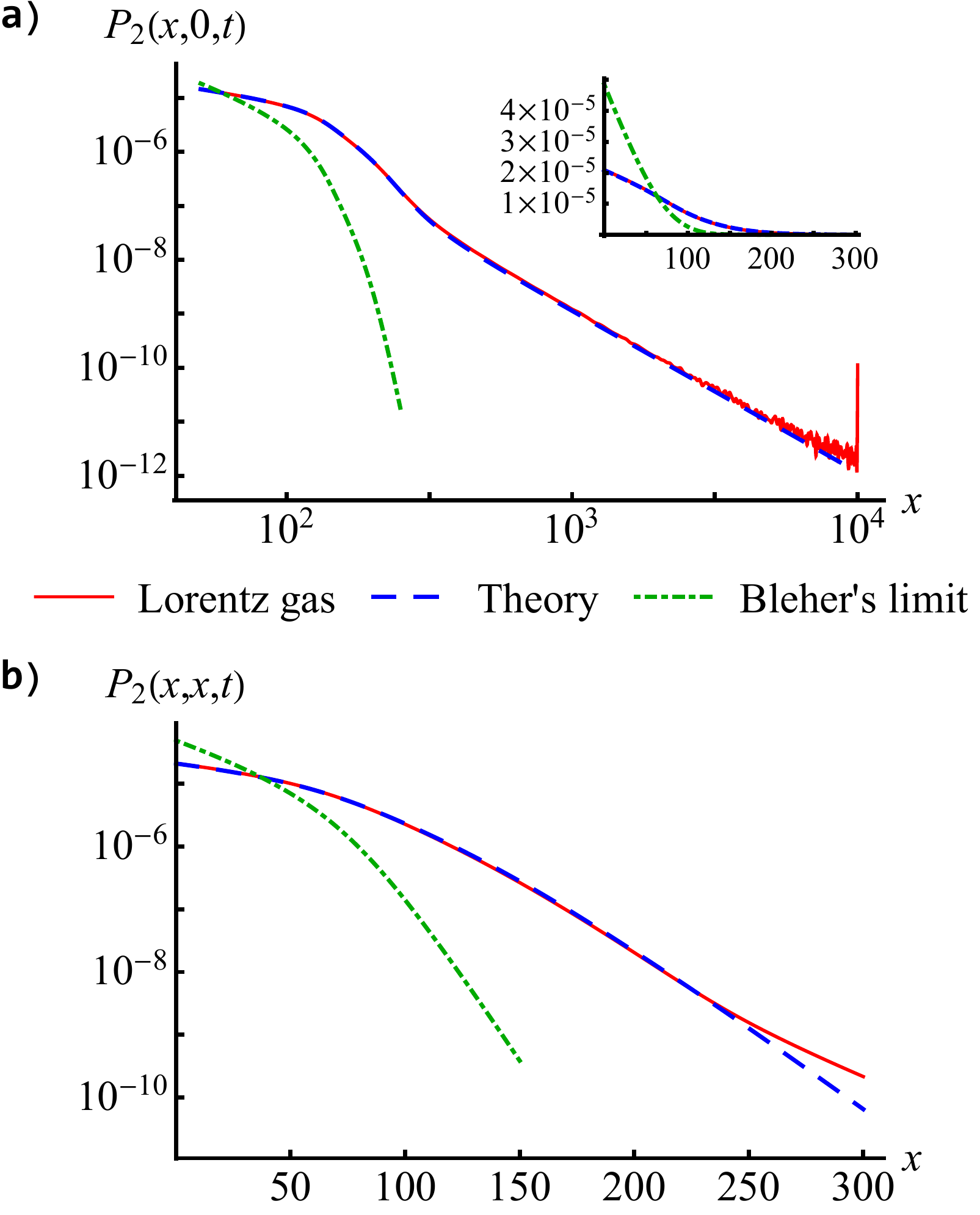}
	\caption{Cross sections of the probability density function of the Lorentz gas with $R=0.4$, Fig.~\ref{LorentzCross3D}, for $y=0$ (a) and $y=x$ (b). The green dot-dashed curve is Bleher's Gaussian limit. Solid red is the Lorentz gas numerical simulations, dashed blue is Eq.~(\ref{equation: position distribution for the gas}) with $q=0$. Deviations in the $x\approx300$ area of the $y=x$ case are caused by the finite number of sampled trajectories $\approx 10^9$.}
	\label{LorentzCross2D}
\end{figure}

To obtain the distribution of traveling times we define a $2d$ cubic lattice of constant $1$ occupied with circular scatterers of radius $R$, such that the center of each circle is located at a grid point, see Fig.~\ref{SetupSketches} (a) and (b). Let the lattice coordinates of each scatterer be $(n,m)$, where $n$ and $m$ are integers. We focus on the origin, and assume that the particle has just collided with the $(0,0)$ scatterer. We define the collision's impact parameter and recoil direction as $b$ and $\beta$ respectively, see Fig.~\ref{ParametersDefinitionsCDF} (a), where the ranges of values for these two parameters are $[-R,R]$ and $[0,2\pi)$, respectively. We denote as $\tau_{n,m}^*(b,\beta,R)$ the time duration until the following collision, and since $V=1$ it is also the distance traveled. Here, we assume that the next scatterer to be collided with is the $(n,m)$ one. The expression obtained for $\tau_{n,m}^*(b,\beta,R)$ is
\begin{align}
\label{equation: tau star definition}
	\tau_{n,m}^*(b,\beta,R) &= n\cos(\beta) + m\sin(\beta) -\sqrt{R^2-b^2} \nonumber \\
	&-\sqrt{ R^2 - \left[ m\cos(\beta) - n\sin(\beta) - b \right]^2} .
\end{align}
We then write the PDF $\psi(\tau)$ as
\begin{align}
\label{equation: psi as sum of integrals}
	&\psi(\tau) = \\
	&\sum_{n,m} \int_{\beta_{n,m}^{\rm min}(R)}^{\beta_{n,m}^{\rm max}(R)} \frac{\text{d}\beta}{2\pi} \int_{b_{n,m}^{\rm min}(\beta,R)}^{b_{n,m}^{\rm max}(\beta,R)} \frac{\text{d}b}{2R} \; \delta\left[ \tau - \tau^*_{n,m}(b,\beta,R) \vphantom{\frac{1}{2}} \right] , \nonumber
\end{align}
where the factors of $1/2\pi \times 1/2R$ are the distributions of $\beta$ and $b$ respectively, which are both uniform due to ergodicity \cite{Bouchaud}. In Eq.~(\ref{equation: psi as sum of integrals}), the summation is carried over all relevant integers (see Fig.~\ref{LorentzSymmetry}), and the integration boundaries (IBs) need to be found. Therefore, we define the discriminant of Eq.~(\ref{equation: tau star definition})
\begin{equation}
\label{equation: tau star discriminant}
	\Delta_{n,m}(b,\beta,R) = R^2 - \left[ m\cos(\beta) - n\sin(\beta) - b \right]^2 .
\end{equation}
Our starting point for obtaining the IBs is to notice that the rooted quantities in Eq.~(\ref{equation: tau star definition}) must be positive. Actually, $R^2-b^2 \ge 0$ means that the particle path indeed intersects the $(0,0)$ circle, which is an initial assumption here, and a positive discriminant in Eq.~(\ref{equation: tau star discriminant}) means that the particle does collide with the $(n,m)$ scatterer. However, in order for the particle to reach the $(n,m)$ scatterer, it must as well not collide with another circle along its path. We therefore use a system of inequalities which are drawn from Eq.~(\ref{equation: tau star discriminant}) to determine $\beta$ and $b$'s IBs. Assuming that $1/\sqrt{8}<R<1/2$, one has a single pair of infinite corridors, and the only scatterers which are reachable to the particle are those with lattice indexes that obey $n=1$ or $m=1$, see the textured colored circles in Fig.~\ref{LorentzSymmetry} (a). Symmetry considerations allow us to break the problem into eight areas composed of three components each, and we choose to focus on $m=1$ and $n\ge0$, see the black rectangle-encircled area in Fig.~\ref{LorentzSymmetry} (a). This area's first component is the nearest neighbor scatterer $(0,1)$. Here, Eq.~(\ref{equation: tau star discriminant}) suggests that $b$ obeys
\begin{equation}
\label{equation: cross NN inequality}
	\Delta_{0,1}(b,\beta,R) \ge 0 .
\end{equation}
The above condition ensures that the particle does collide with the scatterer $(0,1)$ on the next collision. In addition, this domain should be intersected with $|b| \le R$, which ensures us that the particle indeed originated from the $(0,0)$ scatterer. Similarly, for the second component, the next to nearest neighbor scatterer $(1,1)$, the system of inequalities which stems from Eq.~(\ref{equation: tau star discriminant}) reads
\begin{align}
\label{equation: cross NNN inequalities}
	&\Delta_{1,1}(b,\beta,R) \ge 0 , \nonumber \\
	&\Delta_{1,0}(b,\beta,R) \le 0 , \nonumber \\
	&\Delta_{0,1}(b,\beta,R) \le 0 ,
\end{align}
and of course $|b|\le R$ as before. This ensures that the particle collides with the $(1,1)$ scatterer (first condition), but not with the $(1,0)$ or the $(0,1)$ circles (second and third conditions, respectively) that will otherwise block its path. Lastly, the third component in this area is the distant neighbors cluster $n>1$, which is dominated by the inequalities
\begin{align}
\label{equation: cross DN inequalities}
	\Delta_{n,1}(b,\beta,R) &\ge 0 , \nonumber \\
	\Delta_{1,0}(b,\beta,R) &\le 0 , \nonumber \\
	\Delta_{n-1,1}(b,\beta,R) &\le 0 ,
\end{align}
as well as $|b|\le R$. This ensures that the particle collides with the $(n,1)$ scatterer, but avoids from the $(1,0)$ or the $(n-1,1)$ ones, which can block its path. Analyzing these inequalities in appendix \ref{appendix: integration boundaries calculation}, we obtain
\begin{equation}
\label{equation: cross lower beta boundary}
	\beta_{n,1}^{\rm min}(R) = \sin^{-1}\left(\frac{1}{\sqrt{n^2+1}}\right) - \sin^{-1}\left(\frac{2R}{\sqrt{n^2+1}}\right) ,
\end{equation}
for the lower $\beta$ IB,
\begin{equation}
\label{equation: cross upper beta boundary}
	\beta_{n,1}^{\rm max}(R) = \left\{
	\begin{aligned}
		&\pi/2 & n=0 \\
		&\pi/4 & n=1 \\
		&\beta_{n-2,1}^{\rm min}(R) & n>1
	\end{aligned}
	\right. ,
\end{equation}
for the upper $\beta$ IB, and
\begin{align}
\label{equation: cross b boundaries}
	&b_{n,1}^{\rm min}(\beta,R) + R = \\
	&\left\{
	\begin{aligned}
		&\cos(\beta)-n\sin(\beta) & \beta_{n,1}^{\rm min}(R) \le \beta \le \beta_{n,1}^{\rm sep}(R) \\
		&2R-\sin(\beta) & \beta_{n,1}^{\rm sep}(R) \le \beta \le \beta_{n,1}^{\rm max}(R)
	\end{aligned}
	\right. , \nonumber \\
	&b_{n,1}^{\rm max}(\beta,R) + R = \nonumber \\
	&\left\{
	\begin{aligned}
		&2R & \beta_{n,1}^{\rm min}(R) \le \beta \le \beta_{n,1}^{\rm sep}(R) \\
		&\cos(\beta)-(n-1)\sin(\beta) & \beta_{n,1}^{\rm sep}(R) \le \beta \le \beta_{n,1}^{\rm max}(R)
	\end{aligned}
	\right. , \nonumber
\end{align}
for the IBs of $b$, where
\begin{equation}
\label{equation: cross beta separator}
	\beta_{n,1}^{\rm sep}(R) = \left\{
	\begin{aligned}
		&\pi/2 & n=0 \\
		&\beta_{n-1,1}^{\rm min}(R) & n>0
	\end{aligned}
	\right. .
\end{equation}
Equations~(\ref{equation: cross lower beta boundary}-\ref{equation: cross beta separator}) together with Eq.~(\ref{equation: psi as sum of integrals}) provide a full description of $\psi(\tau)$, and alternatively by an integration over $\tau$, a full description of $\text{CDF}(\tau)$. Figure~\ref{LorentzTauCDF} (a) depicts the CDF obtained from these equations for $R=0.4$ with its respective numerical simulations counterpart, where an excellent match can be witnessed. With these equations in mind, we derive an exact expression for $\tau_0$, and numerical values of $\langle\tau\rangle$ and $C_{\psi}$ for $R=0.4$, see appendix \ref{appendix: constants calculation}. For $\tau_0$ we obtain
\begin{equation}
\label{equation: fat tail tau zero}
	\tau_0^2 = \frac{2}{\pi R}(1-2R)^2 .
\end{equation}
This matches the limiting result obtained in Ref.~\cite{Bouchaud} for $R\to1/2$. It is also in perfect match to our previous exact result for $\tau_0$ which we obtained in Ref.~\cite{PRE2018} using a different indirect approach. For the aforementioned numerical values, we get $\langle\tau\rangle\approx 0.62155$ and $C_{\psi}\approx 4.4802\times 10^{-4}$, where the former has a relative error of $0.021\%$ to the known rigorous result
\begin{equation}
\label{equation: tau average exact}
	\langle\tau\rangle = \frac{1-\pi R^2}{2R} ,
\end{equation}
which is mentioned in \cite{Dettmann1}. These values provide excellent results for the numerical simulations of the position's PDF when used as an input for the L\'evy walk approximation, as seen in Figs.~\ref{LorentzCross3D} and \ref{LorentzCross2D}.

\begin{figure}
	\includegraphics[width=1.0\columnwidth]{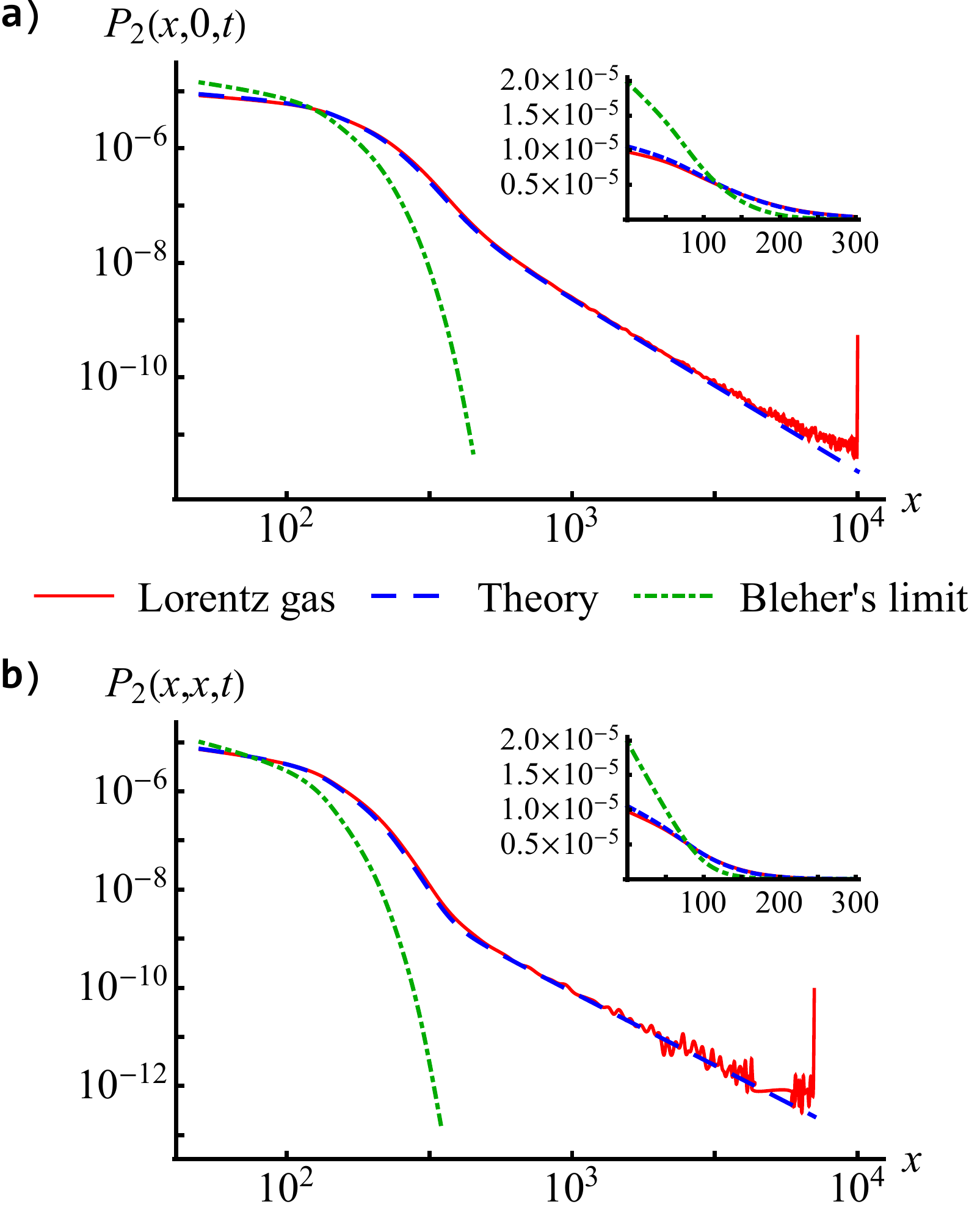}
	\caption{Cross sections of the probability density function of the Lorentz gas with $R=0.3$, Fig.~\ref{LorentzFlag3D}, for $y=0$ (a) and $y=x$ (b). The green dot-dashed curve is Bleher's Gaussian limit. Solid red is the Lorentz gas numerical simulations, dashed blue is Eq.~(\ref{equation: position distribution for the gas}) with $q>0$ which is given by Eq.~(\ref{equation: positive q definition}). The curves match perfectly.}
	\label{LorentzFlag2D}
\end{figure}

Next we compute the case of four open horizons for the Lorentz gas model, see Figs.~\ref{LorentzFlag3D} and \ref{LorentzFlag2D}. Assuming that $\sqrt{20} \le R < 1/\sqrt{8}$, there are two additional open horizons, namely the two main diagonals. It turns out that the formulas we obtained above for the $m=1$ and $n\ge0$ stripe are valid here as well, excluding the $(2,1)$ scatterer. Since $R$ is now smaller, possible trajectories emerge for the diagonal directions, and we choose to focus on movements which end at the $(m+1,m)$ circles where $m\ge1$, see the ellipse-encircled area in Fig.~\ref{LorentzSymmetry} (b). Thus, we need to adjust the $(2,1)$ scatterer's upper IB, as it is now counted in two distinct sets of scatterers. The specifics are detailed in appendix \ref{appendix: integration boundaries calculation}, and we obtain the correction
\begin{equation}
\label{equation: cross upper beta boundary correction}
	\beta_{2,1}^{\rm max}(R) = \sin^{-1}(2R) .
\end{equation}
The diagonal area's inequalities are obtained yet again from Eq.~(\ref{equation: tau star discriminant}). For the diagonal part of $(2,1)$, we have
\begin{equation}
\label{equation: flag NN inequality}
	\Delta_{2,1}(b,\beta,R) \ge 0 .
\end{equation}
This circle is similar to the $(0,1)$ one in the previous case, as there are no possible obstacles for the (diagonal) nearest neighbor. The rest of the IBs are found by analyzing
\begin{align}
\label{equation: flag DN inequalities}
	\Delta_{m+1,m}(b,\beta,R) &\ge 0 , \nonumber \\
	\Delta_{m,m-1}(b,\beta,R) &\le 0 , \nonumber \\
	\Delta_{1,1}(b,\beta,R) &\le 0 ,
\end{align}
in a similar way as was done previously, where $m>1$. This time, the possible scatterers to block the particle's path are $(m,m-1)$ and $(1,1)$. We obtain the following
\begin{align}
\label{equation: flag upper beta boundary}
	\beta_{m+1,m}^{\rm max}(R) = &\sin^{-1}\left(\frac{m}{\sqrt{m^2+(m+1)^2}}\right) \nonumber \\
	+ &\sin^{-1}\left(\frac{2R}{\sqrt{m^2+(m+1)^2}}\right) ,
\end{align}
for the upper $\beta$ IB,
\begin{equation}
\label{equation: flag lower beta boundary}
	\beta_{m+1,m}^{\rm min}(R) = \left\{
	\begin{aligned}
		&\sin^{-1}(2R) & m=1 \\
		&\sin^{-1}(2R) & m=2 \\
		&\beta_{m-1,m-2}^{\rm max}(R) & m>2
	\end{aligned}
	\right. ,
\end{equation}
for the lower $\beta$ IB, and
\begin{align}
\label{equation: flag b boundaries}
	&b_{m+1,m}^{\rm min}(\beta,R) - R = \\
	&\left\{
	\begin{aligned}
		&(m-1)\cos(\beta)-m\sin(\beta) & \beta_{m+1,m}^{\rm min}(R) \le \beta \le \beta_{m+1,m}^{\rm sep}(R) \\
		&-2R & \beta_{m+1,m}^{\rm sep}(R) \le \beta \le \beta_{m+1,m}^{\rm max}(R)
	\end{aligned}
	\right. , \nonumber \\
	&b_{m+1,m}^{\rm max}(\beta,R) - R = \nonumber \\
	&\left\{
	\begin{aligned}
		&\cos(\beta)-\sin(\beta)-2R & \beta_{m+1,m}^{\rm min}(R) \le \beta \le \beta_{m+1,m}^{\rm sep}(R) \\
		&m\cos(\beta)-(m+1)\sin(\beta) & \beta_{m+1,m}^{\rm sep}(R) \le \beta \le \beta_{m+1,m}^{\rm max}(R)
	\end{aligned}
	\right. , \nonumber
\end{align}
for the IBs of $b$, where
\begin{equation}
\label{equation: flag beta separator}
	\beta_{m+1,m}^{\rm sep}(R) = \left\{
	\begin{aligned}
		&\sin^{-1}(2R) & m=1 \\
		&\beta_{m,m-1}^{\rm max}(R) & m>1
	\end{aligned}
	\right. .
\end{equation}
Using Eqs.~(\ref{equation: flag upper beta boundary}-\ref{equation: flag beta separator}), we calculate the $\text{CDF}(\tau)$ for $R=0.3$, which is plotted in Fig.~\ref{LorentzTauCDF} (b), where an excellent match to the simulations can be seen. We derive from these equations an exact expression for $\tau_0$, obtaining
\begin{equation}
\label{equation: fat tail tau zero extended}
	\tau_0^2 = \frac{2}{\pi R}(1-2R)^2 + \frac{\sqrt{2}}{\pi R}\left(1-\sqrt{8}R\right)^2 .
\end{equation}
Equation~(\ref{equation: fat tail tau zero extended}) is used to define $q$, which determines the relevant weight of the velocities' PDF in the L\'evy walk formalism, along respective directions of the infinite corridors. Recalling that $q$ determines the probability of a particle to be at the diagonal corridors, we \textit{define} this parameter as the ratio between the diagonal corridors' contribution for the behavior $\psi(\tau)\sim\tau^{-3}$ and the overall $\tau_0^2$, and find that
\begin{equation}
\label{equation: positive q definition}
	q = \frac{\left(1-\sqrt{8}R\right)^2}{\sqrt{2}\left(1-2R\vphantom{\sqrt{1}}\right)^2 +\left(1-\sqrt{8}R\right)^2} ,
\end{equation}
when $1/\sqrt{20} \le R < 1/\sqrt{8}$. We also get from $\psi(\tau)$ numerical values for $\langle\tau\rangle$ and $C_{\psi}$ for this specific value of $R$, see appendix \ref{appendix: constants calculation}. We find that $\langle\tau\rangle\approx1.1947$, with a relative error of $0.059\%$ to the rigorous result Eq.~(\ref{equation: tau average exact}), and also $C_{\psi}\approx1.5250\times 10^{-2}$. These values provide excellent results for the numerical simulations of the position's PDF when used as an input for the L\'evy walk approximation, as seen in Figs.~\ref{LorentzFlag3D} and \ref{LorentzFlag2D}.

\subsection{Stadium channel model}

\begin{figure}
	\includegraphics[width=1.0\columnwidth]{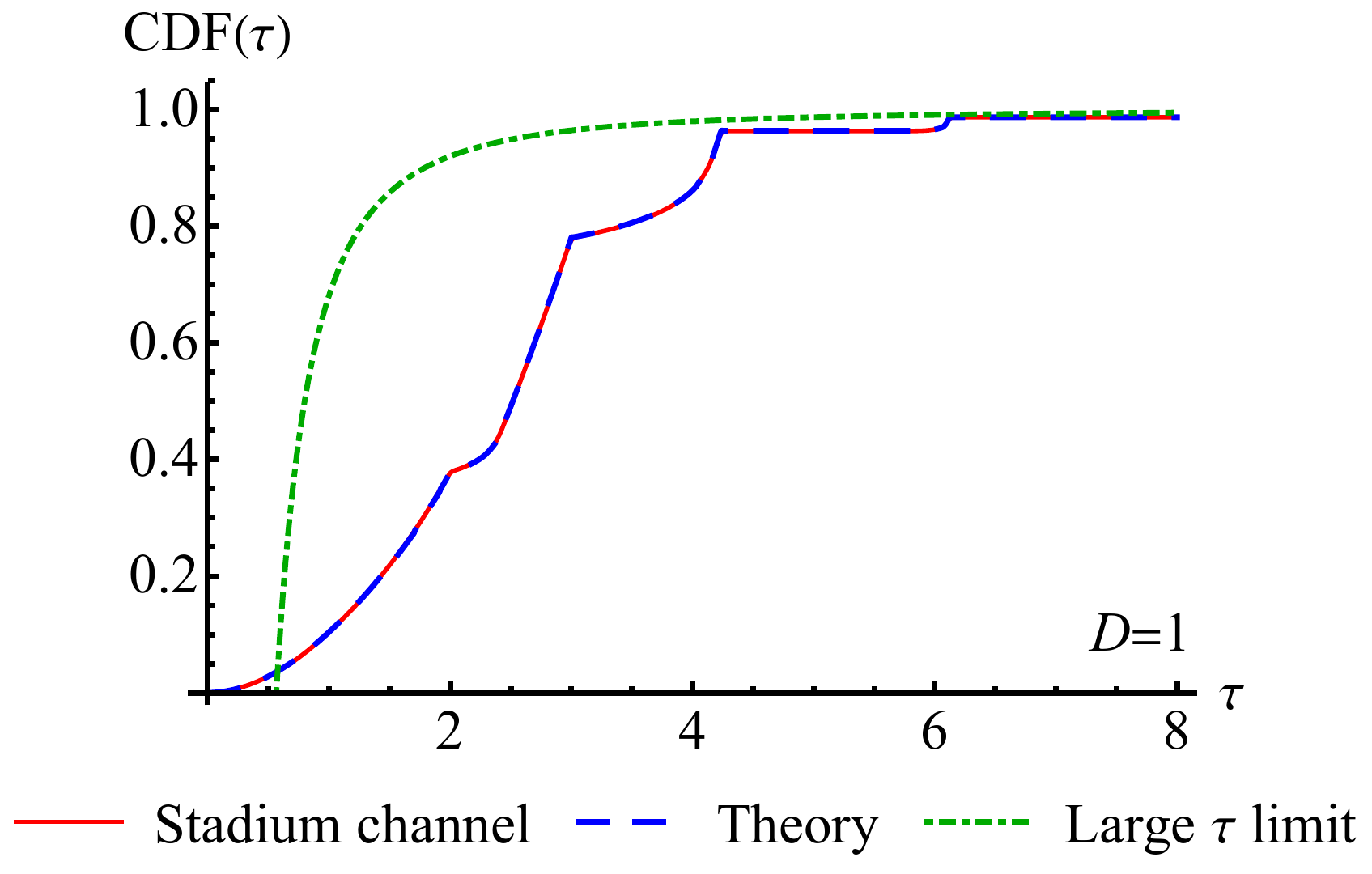}
	\caption{The cumulative distribution function of inter-collision times for the stadium channel model derived in section \ref{section: cdf calculation}, its respective result obtained from the numerical simulations, and the large $\tau$ limit given by $\text{CDF}(\tau)\simeq 1-\tau_0^2/2\tau^2$. The walls of semicircles are distanced $D=1$ from each other, and the radius of a stadium is $1$. The numerical histograms are made of a single long trajectory containing $10^6$ collisions.}
	\label{PipeTauCDF}
\end{figure}

Let us now precisely define the notation used for the stadium channel. As we have a lower and upper boundaries for this pipe structure, we define two parallel one-dimensional straight lattices of constant $2$ which are separated by a distance $D$. These are occupied with circular stadiums of radius $1$, such that the center of each stadium is located at a grid point, see Fig.~\ref{SetupSketches} (c). Let $(2n,m)$ denote the center of a given stadium, where $n$ is an integer and $m$ can take two possible values, $0$ and $D$ [see Fig.~\ref{ParametersDefinitionsCDF} (b)]. We focus on the origin, and assume that the particle has just been scattered from the lower wall's $n=0$ stadium. We define the collision's impact parameter and recoil direction as $a$ and $\alpha$ respectively, see Fig.~\ref{ParametersDefinitionsCDF} (b), where the ranges of values for these two parameters are $[-1,1]$ and $[0,2\pi)$, respectively. We denote as $\tau_{2n,m}^*(a,\alpha)$ the time duration until the following collision, and since $V=1$ it is also the distance traveled. Here, the next stadium to be collided with is $(2n,m)$, where the integer $n$ can be regarded as a semicircle's numbering, while $m$ denotes the top or bottom wall. Notice that the particle is able to reach the stadium of origin, namely $n=m=0$. We obtained for $\tau_{2n,m}^*(a,\alpha)$
\begin{align}
\label{equation: tau star definition pipe}
	\tau_{2n,m}^*(a,\alpha) &= 2n\cos(\alpha) + m\sin(\alpha) + \sqrt{1-a^2} \nonumber \\
	&+\sqrt{1 - \left[ m\cos(\alpha) - 2n\sin(\alpha) - a \right]^2} .
\end{align}
The only semicircles which are reachable to the particle are those located at the upper row, or the origin stadium. Symmetry considerations allow us to break the problem into two areas, and we choose to focus on $n\ge0$, see Fig.~\ref{SetupSketches} (c). We then write the PDF $\psi(\tau)$ as
\begin{align}
\label{equation: psi as sum of integrals pipe}
	&\psi(\tau) \\
	&=4\int_{\alpha_{0,0}^{\rm min}}^{\alpha_{0,0}^{\rm max}} \frac{\text{d}\alpha}{2\pi} \int_{a_{0,0}^{\rm min}(\alpha)}^{a_{0,0}^{\rm max}(\alpha)} \frac{\text{d}a}{2} \; \delta\left[ \tau - \tau^*_{0,0}(a,\alpha) \vphantom{\frac{1}{2}} \right] \nonumber \\
	&+4\sum_{n=0}^{\infty} \int_{\alpha_{2n,D}^{\rm min}}^{\alpha_{2n,D}^{\rm max}} \frac{\text{d}\alpha}{2\pi} \int_{a_{2n,D}^{\rm min}(\alpha)}^{a_{2n,D}^{\rm max}(\alpha)} \frac{\text{d}a}{2} \; \delta\left[ \tau - \tau^*_{2n,D}(a,\alpha) \vphantom{\frac{1}{2}} \right] \nonumber ,
\end{align}
where the factors of $1/2\pi \times 1/2$ are the distributions of $\alpha$ and $a$ respectively. As the corresponding parameters of the Lorentz gas $\beta$ and $b$ are known to be uniform \cite{Bouchaud}, here we \textit{assume} the same for $\alpha$ and $a$. The particle can travel into the $n \le 0$ area, as well as from top to bottom, hence the multiplicative factor of $4$. To obtain the IBs, we use a similar scheme as for the Lorentz gas. However, there is a major difference between the two cases. Due to convexity, there are two possible points of origin/collision for the particle's trajectory. In the Lorentz gas case, there was no need to differentiate between these two options during the calculations, thus we used the discriminant to tell whether a given scatterer was hit/missed. In the stadiums channel case, the discriminant is of no use as the stadiums are semicircles, and the discriminant cannot differentiate between a true stadium and a continuation of its wall to a complete circle. Therefore, here we define the vertical axis coordinate of the origin and target points, $y_0(a,\alpha)$ and $w_{2n,m}(a,\alpha)$ respectively
\begin{align}
\label{equation: y component origin target points}
	&y_0(a,\alpha) = \sqrt{1-a^2}\sin(\alpha) + a\cos(\alpha) , \nonumber \\
	&w_{2n,m}(a,\alpha) = y_0(a,\alpha)+\tau_{2n,m}^*(a,\alpha)\sin(\alpha) ,
\end{align}
and extract the needed inequalities from them instead. This time we demand that $w_{2n,m}(a,\alpha)-m$ will be non-negative for the upper stadiums, and non-positive for the origin stadium. This replaces the demand of a positive discriminant in the Lorentz gas case. We also demand that $y_0(a,\alpha)$ will be non-positive, which is analogous to $|b|\le R$ in the Lorentz gas case. The first component of the chosen area is the origin semicircle, for which Eq.~(\ref{equation: y component origin target points}) dictates that $a$ obeys
\begin{equation}
\label{equation: pipe NN inequalities}
	y_0(a,\alpha) \le 0 , \quad w_0(a,\alpha,0) \le 0 .
\end{equation}
The upper IB of $\alpha$ is set to $\pi/2$, and the lower to $-\pi/2$, which is possible due to symmetry. The second component is the $n=0$ upper stadium, for which we have
\begin{equation}
\label{equation: pipe NNN inequalities}
	y_0(a,\alpha) \le 0 , \quad w_{0,D}(a,\alpha) \ge D .
\end{equation}
The upper IB of $\alpha$ is set again to $\pi/2$ for symmetry considerations, and the lower is found by analyzing Eq.~(\ref{equation: pipe NNN inequalities}). Lastly, in order for the particle to reach the $n>0$ upper semicircles, it must not collide with another stadium wall along its path. For this third component, the conditions are
\begin{align}
\label{equation: pipe DN inequalities}
	y_0(a,\alpha) &\le 0 , &w_{2n,D}(a,\alpha) &\ge D , \nonumber \\
	w_{2n-2,D}(a,\alpha) &\le D , &w_{0,0}(a,\alpha) &\ge 0 .
\end{align}
The upper row in Eq.~(\ref{equation: pipe DN inequalities}) ensures that the particle originated from and arrived to the correct points, while the lower row prevents the top $2n-2$ and bottom $n=0$ semicircles from blocking the particle's path. Analyzing these inequalities in appendix \ref{appendix: integration boundaries calculation}, we obtain for the origin stadium as the target, $n=m=0$
\begin{align}
\label{equation: pipe origin as target boundaries}
	&-1 \le a \le -\sin(\alpha) , &0 \le \alpha \le \frac{\pi}{2} \nonumber \\
	&-1 \le a \le \sin(\alpha) , &-\frac{\pi}{2} \le \alpha \le 0 ,
\end{align}
and for the top $n\ge 0$ and $m=D$ row of stadiums as targets
\begin{equation}
\label{equation: pipe lower alpha boundary}
	\alpha_{2n,D}^{\rm min} = \tan^{-1}\left(\frac{D}{2n+2}\right) ,
\end{equation}
for the lower $\alpha$ IB,
\begin{equation}
\label{equation: pipe upper alpha boundary}
	\alpha_{2n,D}^{\rm max} = \left\{
	\begin{aligned}
		&\pi/2 & n=0,1 \\
		&\alpha_{2n-4,D}^{\rm min} & n>1
	\end{aligned}
	\right. ,
\end{equation}
for the upper $\alpha$ IB, and
\begin{align}
\label{equation: pipe a boundaries}
	&a_{2n,D}^{\rm min}(\alpha) = \\
	&\left\{
	\begin{aligned}
		&D\cos(\alpha)-(2n+1)\sin(\alpha) & \alpha_{2n,D}^{\rm min} \le \alpha \le \alpha_{2n,D}^{\rm sep} \\
		&-\sin(\alpha) & \alpha_{2n,D}^{\rm sep} \le \alpha \le \alpha_{2n,D}^{\rm max}
	\end{aligned}
	\right. , \nonumber \\
	&a_{2n,D}^{\rm max}(\alpha) = \nonumber \\
	&\left\{
	\begin{aligned}
		&\sin(\alpha) & \alpha_{2n,D}^{\rm min} \le \alpha \le \alpha_{2n,D}^{\rm sep} \\
		&D\cos(\alpha)-(2n-1)\sin(\alpha) & \alpha_{2n,D}^{\rm sep} \le \alpha \le \alpha_{2n,D}^{\rm max}
	\end{aligned}
	\right. , \nonumber
\end{align}
for the IBs of $a$, where
\begin{equation}
\label{equation: pipe alpha separator}
	\alpha_{2n,D}^{\rm sep} = \left\{
	\begin{aligned}
		&\pi/2 & n=0 \\
		&\alpha_{2n-2,D}^{\rm min} & n>0
	\end{aligned}
	\right. .
\end{equation}
Figure~\ref{PipeTauCDF} depicts the CDF obtained from Eqs.~(\ref{equation: pipe lower alpha boundary}-\ref{equation: pipe alpha separator}) for $D=1$ with its respective numerical simulations counterpart, where an excellent match can be witnessed. We also use these to derive an exact expression for $\tau_0$, and the numerical values of $\langle\tau\rangle$ and $C_{\psi}$ for $D=1$, see appendix \ref{appendix: constants calculation}. For $\tau_0$, we obtain
\begin{equation}
\label{equation: fat tail tau zero pipe}
	\tau_0 = \sqrt{\frac{2}{\pi}}D .
\end{equation}
We also find $\langle\tau\rangle\approx 2.57016$, with a relative error of $0.005\%$ to the simulation result $\langle\tau\rangle\approx 2.57031$, and $C_{\psi}\approx 1.0903\times 10^{-5}$. However, these values do not provide a correct description for the L\'evy walk approximation, due to the renewal assumption being nullified by strong temporal correlations discussed below. Nonetheless, one can fit the L\'evy walk approximation to the simulations data using a two parameters fit ($C_{\psi}^2\langle\tau\rangle$ and $\langle\tau\rangle/\tau_0^2$), thus find ``effective" constants. These turn out to describe the problem well, as seen in Fig.~\ref{PipePositionPDF} (a). We have verified that the values obtained for the constants by fitting do not change over time, see Fig.~\ref{PipePositionPDF} (b).

We would like to direct the reader's attention to the different geometry of the Lorentz gas CDF and the stadium channel CDF. Both CDFs have qualitatively different shapes, being convex/concave for the Lorentz gas/stadium channel, see Figs.~\ref{LorentzTauCDF}/\ref{PipeTauCDF} respectively. This might be related to the scatterers' shape in the two models, which is convex/concave for the Lorentz gas/stadium channel, respectively. We believe this phenomenon is general as the geometry of the scattering centers is clearly embedded in this basic distribution, however we leave this intriguing point for a future work.

\begin{figure*}
	\includegraphics[width=1.0\textwidth]{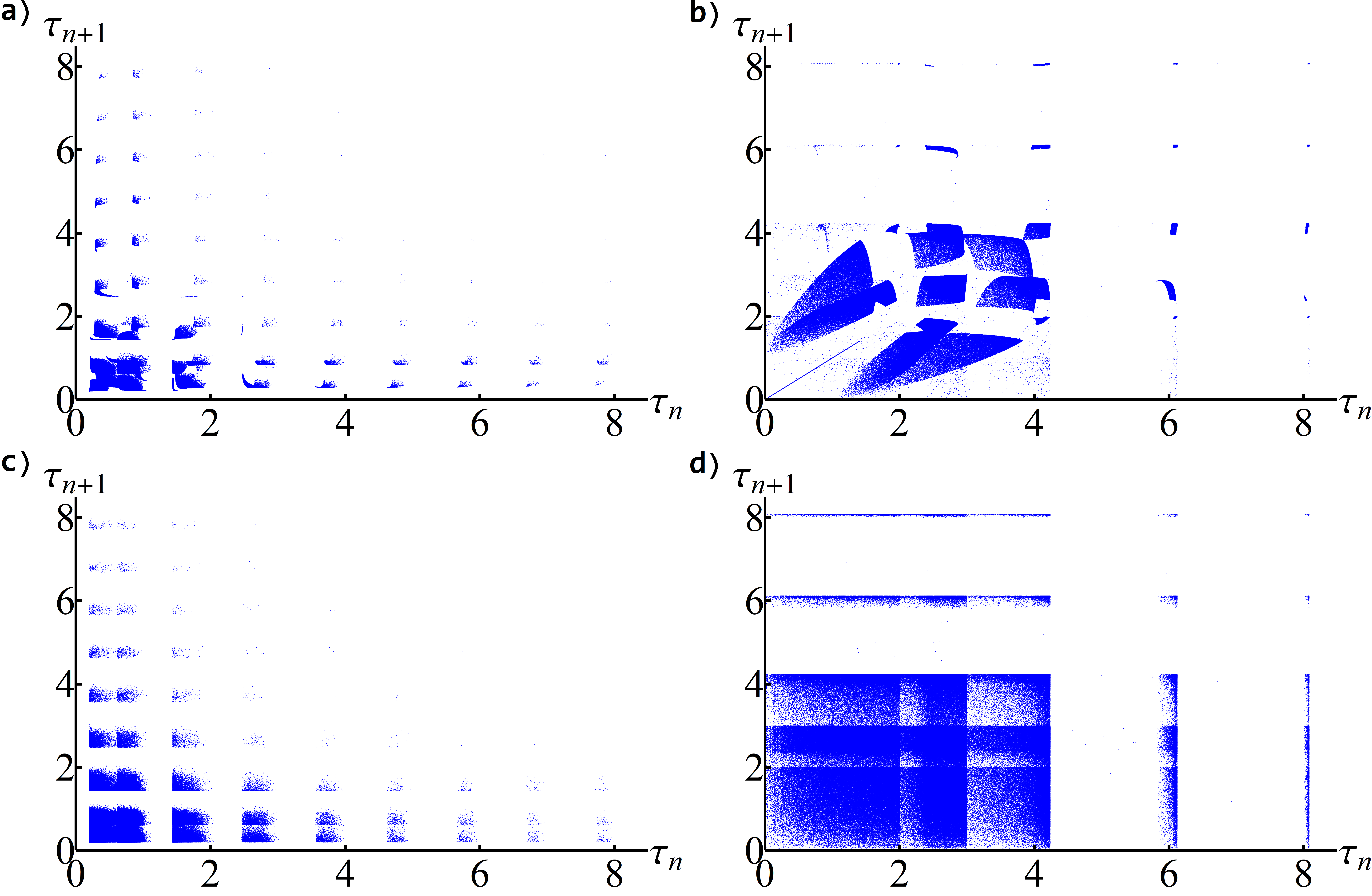}
	\caption{Correlations of the traveling times between collisions. Seen are points of the form $(\tau_n,\tau_{n+1})$, where $\tau_n$ is the $n$th flight duration. (a) and (b) depict numerical realizations of the Lorentz gas model with two open horizons ($R=0.4$) and the stadium channel model ($D=1$), respectively. (c) and (d) present points drawn from the inter-collision times' CDF obtained analytically for the Lorentz gas with two infinite corridors Eq.~(\ref{equation: psi as sum of integrals}) ($R=0.4$) and for the stadium channel Eq.~(\ref{equation: psi as sum of integrals pipe}) ($D=1$), respectively. Each plot consists out of approximately $10^6$ points. The Lorentz gas displays strong similarity between repeated draws and the simulation data, meaning that the renewal condition is indeed fulfilled. However, for the stadium channel patterns are substantially different, which means that here the condition fails. See additional discussion in section \ref{section: discussion}.}
	\label{TemporalCorrelations}
\end{figure*}

\begin{figure}
	\includegraphics[width=1.0\columnwidth]{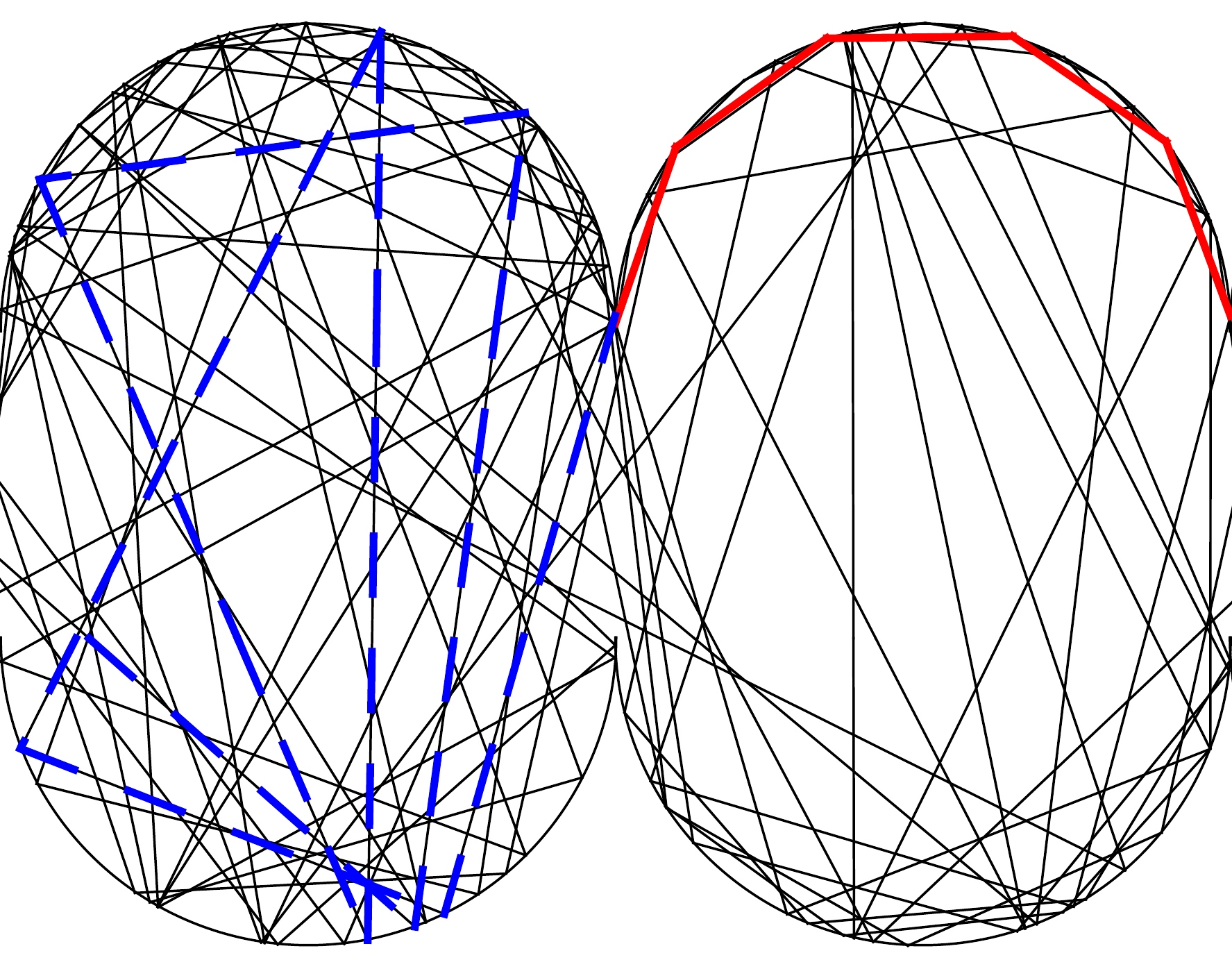}
	\caption{A piece of trajectory for the stadium channel model, with colored correlative sequences. Right cell - ``gallery whispering paths", which are responsible for the $\tau_{n+1}=\tau_n$ line in Fig~\ref{TemporalCorrelations} (solid thick red). Left cell - up and down repetitive movements termed ``periodic orbits", responsible for the center rectangle-shaped area in Fig~\ref{TemporalCorrelations} (dashed thick blue) discussed in main text. These types of ``traps" hold the particle in a localized area in the phase space \cite{Chernov2}, thus invalidating renewals.}
	\label{PipeTrajectory}
\end{figure}

\section{Discussion and summary}
\label{section: discussion}

Returning to section \ref{section: simplified case of lambert}, we address an issue we previously disregarded with the Lambert scaling approach. The reader may have noticed that the choice of the scaling function $\Omega(t)$ is not unique, but can be determined up to a constant. More accurately, one may choose to separate the logarithmic term into two at an arbitrary point, as one can always write
\begin{equation}
\label{equation: lambert scaling problem}
	\ln\left[\frac{2C_f^2\bar{k}^2}{N\Omega(N)}\right] = \ln\left[\frac{2C_f^2\eta}{N\Omega(N)}\right] + \ln\left(\frac{\bar{k}^2}{\eta}\right) .
\end{equation}
Recall that we used the first term of Eq.~(\ref{equation: lambert scaling problem})'s right hand side with $\eta=1$ to derive the Lambert scaling, see Eq.~(\ref{equation: lambert scaling choice}), while expanding the $\bar{k}^2$-containing exponential term, see Eq.~(\ref{equation: characteristic function expansion 2}). Here $\eta$ is a free parameter which cannot be determined uniquely by the aforementioned steps alone. We found that taking $\eta=1$ produces good results for $P(x,N)$ and alternatively for $P_d(\boldsymbol{r},t)$, see Figs~\ref{PipePositionPDF}, \ref{PDFofIID}, \ref{LorentzCross2D}, and \ref{LorentzFlag2D}. Of course, if one sums the complete asymptotic series Eq.~(\ref{equation: pdf for simple case}) $\eta$ vanishes, but then the result diverges.

Looking back at Figs.~\ref{LorentzCross3D} and \ref{LorentzFlag3D}, the reader may notice that the shape at $|\boldsymbol{r}| = r \sim t$ of the analytical results mismatches the simulations'. This can be explained via our scaling assumption, $u\sim k^2\ln(k)$. It suggests that $k\gg u$, namely our approximation is for displacement that obey $r \ll t$. Nonetheless, it holds well at the distribution's infinite corridors, $y=0$ and $x=0$ for two open horizons and additionally $y=x$ and $y=-x$ for four. Utilizing the recently gaining attention infinite covariant density \cite{ICD} could probably supply one with tools to approximate edge phenomena such as this.

Finally, we address the reader to interesting correlation patterns of the stadium channel model, seen in Fig.~\ref{TemporalCorrelations} (b). We plot points of consecutive traveling times $(\tau_{n+1},\tau_n)$, for the Lorentz gas and the stadium channel, obtained from numerical samplings and analytical results. Both models exhibit a phenomenon of clear pointless areas on the graph, which is caused by the plateaus in the CDFs Figs.~\ref{LorentzTauCDF} and \ref{PipeTauCDF}, which in turn correspond to a vanishing PDFs $\psi(\tau)$. Indeed, due to the discrete nature of the scattering centers, there exist certain durations of travel that are not possible (in a disordered system this non-analytical behavior would vanish). Technically speaking, large traveling times become semi-discrete, which is due to small parameter spaces $\{b,\beta\}$ [see Eqs.~(\ref{equation: cross lower beta boundary}-\ref{equation: cross b boundaries})] and $\{a,\alpha\}$ [see Eqs.~(\ref{equation: pipe lower alpha boundary}-\ref{equation: pipe a boundaries})] which support these long trajectories. Interesting patterns emerge from roughly $\tau_n\le4.25$ for the stadium channel simulations, see Fig.~\ref{TemporalCorrelations} (b). These patterns suggest a highly correlative system. Indeed, when no correlations are present, one expects to find square-shaped patterns that correspond to independence of the axes, e.g. as seen for the Lorenz gas case (both analytical and numerical). However, the stadium channel's simulations reveal a complete opposite. Take for example the line $\tau_{n+1}=\tau_n$, extending roughly up to $\tau_n\approx1.5$. Examining the raw data, we found repeated instances of the form $\tau_1,\tau_1,\tau_1,...$, where the number of elements depends on the size of $\tau_1$. Each instance always summed to a value of $\le \pi/2$, the length of a semi-circle arc, see Fig.~\ref{PipeTrajectory} (right cell). In these cases, a particle arrives at an almost vertical angle to one end of an arc and then propagates along it, similarly to gallery whispering modes of wave propagation. Other patterns correspond to different repeated instances, e.g. a particle performing ``periodic orbits" [see Fig.~\ref{PipeTrajectory} (left cell)] will have $\tau_n\approx2+D=3$, which fits the middle rectangle. We computed the correlation functions for both models, and found that the Lorentz gas one decay much faster than the stadium channel one (not shown). Indeed, bounds on the temporal decay of the CDFs were obtained, showing that the Lorentz gas one decays with time in a stretched-exponential form \cite{Chernov1}, while the stadium channel one decays in a polynomial form \cite{Chernov2}. Thus, it is of no surprise that the renewal condition does not hold for the stadium channel.

In conclusion, we presented a geometrical method which yields the CDF of traveling times between collision events for billiard systems. We implemented this method for the Lorentz gas with infinite horizon and the stadium channel models. The approach is based on a shadowing effect (as some scatterers cannot be reached), on symmetry, and also on the uniformity of collision parameters (see Figs.~\ref{ParametersDefinitionsCDF} and \ref{LorentzSymmetry}). Our analytical and numerical results coincide. The distribution of waiting times has two main features. The first is a power-law decay Eq.~(\ref{equation: fat tail}), and the second are its non-analytical features (see Figs.~\ref{LorentzTauCDF} and \ref{PipeTauCDF}). The former is due to the infinite horizon and the latter is obviously related to the periodic array of scattering centers. For example in the infinite horizon Lorentz gas, after a collision event the traveling time of the tracer particle to the next target must be larger or equal to the distance between two adjutant scattering centers (in units of unity velocity). This and other dead times implies the observed non-analytical behavior of the traveling times' distribution. Note that for the channel model arbitrarily short times are possible, due to a ``gallery whispering paths" effect, see Fig.~\ref{PipeTrajectory} (right cell). These exhibit a very correlated motion, as a short time interval between collision events is most likely to follow by another short interval of same size, see Fig~\ref{TemporalCorrelations} (b).

This leads to the second theme of our work: can one use the distribution of time intervals between collision events to predict the time dependent probability to find particles at a given position? The answer is system dependent. The technique to make this step is based on the L\'evy walk model, in dimension one for the channel and two for the Lorentz gas. This model makes the assumption of renewal, and we found it works well for the Lorentz gas and not at all for the channel. A tool to check the renewal hypothesis, from an analysis of the paths, is a correlation plot of consecutive waiting times, see Fig.~\ref{TemporalCorrelations} (b) and (d), which clearly points out the strong correlations for the channel model. However, even the failure of renewal theory does not imply the complete breakdown of the L\'evy walk scheme. In this case we introduced effective (or re-normalized) parameters in the L\'evy walk scheme obtained by fitting, yielding predictions that are still very useful. In fact there are general trends in the position's distribution that are universal, and nicely predicted by the L\'evy walk. These include the fat tail of the spreading packet, sharp cutoffs of the density at $|\boldsymbol{r}| = V t$, the Kummer corrections to the Gaussian (which are certainly not small on any reasonable time scale), and the Lambert scaling. The latter is very important since it allowed us to compare finite time simulations with our theory, while the asymptotic Gaussian form (which exist for the Lorentz gas) is not witnessed due to super-slow convergence problems (see Figs.~\ref{LorentzCross2D} and \ref{LorentzFlag2D}). One way to understand this behavior is to realize that the billiard systems are operating at a transition point between L\'evy and Gauss statistics. Because of the exponent $-3$ in Eq.~(\ref{equation: fat tail}), the system is essentially behaving as if is critical in the sense of very sluggish convergence. Roughly speaking and for finite times, the packet of particles' tails exhibit L\'evy behavior (a power-law with cutoff), while the center part is Gaussian. Already in the problem of IID RVs summation, section \ref{section: simplified case of lambert}, we encountered a critical slowing down at this borderline case, solved by departing from the $\sqrt{N\ln(N)}$ scaling and replacing it with the Lambert approach.

To map the problem to a L\'evy walk one needs to model the distribution of velocities $F_d(v)$. For the channel model this is rather easy, as the model is one dimensional and from symmetry we use a velocity which is either $+V$ or $-V$ with equal probability. For the Lorentz gas, a more careful analysis is needed. As we decrease the size of scatterers, we open more infinite corridors of motion. At first we have four open horizons and this leads to a cross-like shape of the spreading packet, see Fig.~\ref{LorentzCross3D}. Here, the velocity distribution in the L\'evy walk scheme has a simple structure as the four directions are clearly identical from symmetry. However, when $R$ is made slightly smaller than $1/\sqrt{8}$, we open a new channel but only slightly, meaning that the effective velocity in these directions is statistically reduced compared to the original four corridors (note that we refer to the distribution of velocities in the L\'evy walk, and not to the microscopic velocities of the Lorenz gas). The resulting effect is the creation of a British flag-like type of structure for the packet's distribution, see Fig.~\ref{LorentzFlag3D}. Thus, we observe a transition from one geometry to another as we vary $R$. Theoretically, this means that we assign different statistical weights to each group of horizons as in Eq.~(\ref{equation: position distribution for the gas}), which is made possible with geometrical considerations leading to the important parameter $q$ in Eq.~(\ref{equation: positive q definition}).

\begin{acknowledgments}

LZ would like to thank Jakub {\'S}l\k{e}zak for helpful conversations. EB thanks Itzhak Fouxon for valuable discussions on the L\'evy walk model. After this paper was completed a related work on the L\'evy walk and Lorentz gas was presented in \cite{Itzhak}. The financial support of Israel Science Foundation's grant $1898/17$ is acknowledged (LZ and EB).

\end{acknowledgments}

\appendix

\begin{widetext}

\section{Leading behavior of $\tilde{f}(k\to 0)$}
\label{appendix: derivation small k}

We assume that $f(\chi)$ possesses the asymptotic behavior
\begin{equation}
\label{equation: f asymptotic chi}
	\lim_{\chi\to\pm\infty} f(\chi)|\chi|^{1+\nu} = \chi_0^{\nu} ,
\end{equation}
where $\nu>0$ and $\chi_0>0$ are real numbers. The Fourier transform of $f(\chi)$ is defined as
\begin{equation}
\label{equation: fourier transform definition}
	\tilde{f}(k) = \int_{-\infty}^{\infty} \text{d}\chi \, f(\chi) e^{ik\chi} .
\end{equation}
Due to the evenness of $f(\chi)$, $\tilde{f}(k)$ is even and real. Therefore, throughout the following sections we assume that $k\to 0^+$, and use parity to find $\tilde{f}(k\to 0^-)$. As we show below, given a positive integer $\bar{n}$, Eqs.~(\ref{equation: f asymptotic chi}-\ref{equation: fourier transform definition}) lead to
\begin{equation}
\label{equation: f asymptotic k}
	\tilde{f}(k) \simeq \sum_{n=0}^{\bar{n}-1} \frac{(ik)^n}{n!}\left<\chi^n\right> + \left\{
	\begin{aligned}
		& 2\Gamma(-\nu)\cos\left(\frac{\pi\nu}{2}\right)|\chi_0k|^{\nu} & \bar{n}-1<\nu<\bar{n} \\
		& \frac{\pi}{\bar{n}!} (-1)^{\frac{\bar{n}+1}{2}} |\chi_0k|^{\bar{n}} & \nu=\text{odd }\bar{n} \\
		& \frac{1}{\bar{n}!} (-1)^{\frac{\bar{n}+2}{2}} (\chi_0k)^{\bar{n}} \ln\left(C_{\bar{n}}^2 \chi_0^2 k^2\right) & \nu=\text{even }\bar{n}
	\end{aligned}
	\right. ,
\end{equation}
where $C_{\bar{n}}\left[f(\chi)\right]$ is defined in the corresponding section. Using Eqs.~(\ref{equation: f asymptotic chi}) and (\ref{equation: fourier transform definition}), we extract all of the converging moments out of the Fourier transform integral
\begin{equation}
\label{equation: fourier moments extracted}
	\tilde{f}(k) = \sum_{n=0}^{\bar{n}-1} \frac{(ik)^n}{n!}\left<\chi^n\right> + \int_{-\infty}^{\infty} \text{d}\chi \, f(\chi) \left[ e^{ik\chi} - \sum_{n=0}^{\bar{n}-1} \frac{(ik\chi)^n}{n!}\right] ,
\end{equation}
and consider each of the cases in Eq.~(\ref{equation: f asymptotic k}) separately.

\subsection{A non-integer $\nu$}

Let us assume that $\bar{n}-1<\nu<\bar{n}$. In order to find the leading behavior of the second term of Eq.~(\ref{equation: fourier moments extracted}), we consider the limit
\begin{equation}
	l_0 = \lim_{k \to 0^+} \frac{1}{k^{\nu}}\int_{-\infty}^{\infty} \text{d}\chi \, f(\chi) \left[ e^{ik\chi} - \sum_{n=0}^{\bar{n}-1} \frac{(ik\chi)^n}{n!}\right] .
\end{equation}
Using L'Hospital's rule $\bar{n}$ times yields
\begin{equation}
	l_0 = \lim_{k \to 0^+} \frac{\Gamma(-\nu)(-i)^{\bar{n}}}{\Gamma(-\nu+\bar{n})} k^{\bar{n}-\nu} \int_{-\infty}^{\infty} \text{d}\chi \, f(\chi) \chi^{\bar{n}} e^{ik\chi} ,
\end{equation}
where $\Gamma(\cdots)$ is the Gamma function. We use an $\eta=k\chi$ variable change
\begin{equation}
	l_0 = \frac{\Gamma(-\nu)(-i)^{\bar{n}}}{\Gamma(-\nu+\bar{n})} \lim_{k \to 0^+} \int_{-\infty}^{\infty} \text{d}\eta \, e^{i\eta} \eta^{\bar{n}} \frac{f(\eta/k)}{k^{\nu+1}} .
\end{equation}
We now switch the order of limit and integration while using the asymptotic behavior (\ref{equation: f asymptotic chi})
\begin{equation}
\label{equation: f non integer limit}
	l_0 = \frac{\chi_0^{\nu}\Gamma(-\nu)(-i)^{\bar{n}}}{\Gamma(-\nu+\bar{n})}\int_{-\infty}^{\infty} \text{d}\eta \frac{e^{i\eta}\eta^{\bar{n}}}{|\eta|^{\nu+1}} = 2\chi_0^{\nu}\Gamma(-\nu)\cos\left(\frac{\pi\nu}{2}\right) ,
\end{equation}
which proves the top row of Eq.~(\ref{equation: f asymptotic k}).

\subsection{An integer $\nu$}

Let us assume that $\nu=\bar{n}$, where $\bar{n}$ is even. In order to find the leading behavior of the second term of Eq.~(\ref{equation: fourier moments extracted}), we consider the limit
\begin{equation}
	l_1 = \lim_{k \to 0^+} \frac{1}{k^{\bar{n}}\ln(k)} \int_{-\infty}^{\infty} \text{d}\chi \, f(\chi) \left[ e^{ik\chi} - \sum_{n=0}^{\bar{n}-1} \frac{(ik\chi)^n}{n!}\right] .
\end{equation}
Using L'Hospital's rule $\bar{n}+1$ times produces
\begin{equation}
	l_1 = \frac{i^{\bar{n}+1}}{\bar{n}!} \lim_{k \to 0^+} k \int_{-\infty}^{\infty} \text{d}\chi \, f(\chi) \chi^{\bar{n}+1} e^{ik\chi} .
\end{equation}
Changing the integration variable to $\eta=k\chi$ gives
\begin{equation}
	l_1 = \frac{i^{\bar{n}+1}}{\bar{n}!} \lim_{k \to 0^+} \int_{-\infty}^{\infty} \text{d}\eta \, e^{i\eta} f\left(\frac{\eta}{k}\right) \left(\frac{\eta}{k}\right)^{\bar{n}+1} .
\end{equation}
After switching the order of limit and integration, the integral exists as a Cauchy principal value, and we find
\begin{equation}
	l_1 = \chi_0^{\bar{n}}\frac{i^{\bar{n}+1}}{\bar{n}!} \, \text{P.V.} \int_{-\infty}^{\infty} \text{d}\eta \, e^{i\eta} \, \text{sign}(\eta) = - \frac{2}{\bar{n}!} (-1)^{\frac{\bar{n}}{2}}\chi_0^{\bar{n}} .
\end{equation}
If $\bar{n}$ is odd, we return to Eq.~(\ref{equation: f non integer limit}), and take the limit of $\nu\to\bar{n}$, where $\bar{n}$ is odd. We obtain
\begin{equation}
	l_1 = \lim_{\nu\to\bar{n}} 2\chi_0^{\nu}\Gamma(-\nu)\cos\left(\frac{\pi\nu}{2}\right) = \frac{\pi}{\bar{n}!} (-1)^{\frac{\bar{n}+1}{2}}\chi_0^{\bar{n}} .
\end{equation}
To compute the next order correction for the case of an even $\bar{n}$, we calculate the following limit
\begin{equation}
	l_2 = \lim_{k \to 0^+} \frac{1}{k^{\bar{n}}} \left\{ \int_{-\infty}^{\infty} \text{d}\chi \, f(\chi) \left[ e^{ik\chi} - \sum_{n=0}^{\bar{n}-1} \frac{(ik\chi)^n}{n!}\right] + \frac{2}{\bar{n}!} (-1)^{\bar{n}/2} (\chi_0k)^{\bar{n}}\ln(k) \right\} .
\end{equation}
Using L'Hospital's rule $\bar{n}$ times results with
\begin{equation}
\label{equation: f integer limit 1}
	l_2 = \frac{(-1)^{\bar{n}/2}}{\bar{n}!} \lim_{k \to 0^+} \left\{ \int_{-\infty}^{\infty} \text{d}\chi \, f(\chi) \chi^{\bar{n}} e^{ik\chi} + 2\chi_0^{\bar{n}} \left[ \ln(k) + H_{\bar{n}} \vphantom{\frac{1}{1}} \right] \right\},
\end{equation}
where $H_{\bar{n}}=\sum_{n=1}^{\bar{n}} \frac{1}{n}$ is the $\bar{n}$th harmonic number. Since $\bar{n}$ is even, the integral in Eq.~(\ref{equation: f integer limit 1}) can be adjusted to the domain $[0,\infty)$, with $\exp(ik\chi)\to\cos(k\chi)$. We split the adjusted integral at $\chi=\chi_0$:
\begin{equation}
\label{equation: f integer limit 2}
	l_2 = \frac{2}{\bar{n}!}(-1)^{\bar{n}/2} \lim_{k \to 0^+} \left\{\int_0^{\chi_0} \text{d}\chi \, f(\chi) \chi^{\bar{n}} \cos(k\chi) + \int_{\chi_0}^{\infty} \text{d}\chi \, f(\chi) \chi^{\bar{n}} \cos(k\chi) + \chi_0^{\bar{n}} \left[ \ln(k) + H_{\bar{n}} \vphantom{\frac{1}{1}} \right] \right\} .
\end{equation}
We now add and subtract a $\chi_0^{\bar{n}}/\chi^{\bar{n}+1}$ term from $f(\chi)$ in the second integral in Eq.~(\ref{equation: f integer limit 2}):
\begin{align}
\label{equation: f integer limit 3}
	l_2 &= \frac{2}{\bar{n}!}(-1)^{\bar{n}/2} \\
	&\times \lim_{k \to 0^+} \left\{ \int_0^{\chi_0} \text{d}\chi \, f(\chi) \chi^{\bar{n}} \cos(k\chi) + \int_{\chi_0}^{\infty} \text{d}\chi \left[ f(\chi) - \frac{\chi_0^{\bar{n}}}{\chi^{\bar{n}+1}} \right] \chi^{\bar{n}} \cos(k\chi) + \int_{\chi_0}^{\infty} \text{d}\chi \frac{\chi_0^{\bar{n}}}{\chi} \cos(k\chi) + \chi_0^{\bar{n}} \left[ \ln(k) + H_{\bar{n}} \vphantom{\frac{1}{1}} \right] \right\} . \nonumber
\end{align}
Note that due to the asymptotics (\ref{equation: f asymptotic chi}), the middle integral is finite when $k \to 0^+$. Eq.~(\ref{equation: f integer limit 3})'s right integral can be computed explicitly, after which the limit can be evaluated. Finally, we find
\begin{equation}
	l_2 = \frac{2}{\bar{n}!}(-1)^{\bar{n}/2} \left\{ \int_0^{\chi_0} \text{d}\chi \, f(\chi) \chi^{\bar{n}} + \int_{\chi_0}^{\infty} \text{d}\chi \left[ f(\chi) - \frac{\chi_0^{\bar{n}}}{\chi^{\bar{n}+1}} \right] \chi^{\bar{n}} + \chi_0^{\bar{n}} \left[ H_{\bar{n}} - \gamma - \ln(\chi_0) \vphantom{\frac{1}{1}} \right] \right\} ,
\end{equation}
where $\gamma \approx 0.5772$ is Euler's constant. After some algebra we obtain
\begin{equation}
	l_2 = -\chi_0^{\bar{n}}\frac{2}{\bar{n}!} (-1)^{\bar{n}/2} \ln(C_{\bar{n}}\chi_0) ,
\end{equation}
where
\begin{equation}
\label{equation: Cf general}
	C_{\bar{n}}\left[f(\chi)\right] = \exp \left\{ \gamma - H_{\bar{n}} - \int_{0}^{\chi_0} \text{d}\chi \, f(\chi) \left(\frac{\chi}{\chi_0}\right)^{\bar{n}} - \int_{\chi_0}^{\infty} \text{d}\chi \left[ f(\chi)\left(\frac{\chi}{\chi_0}\right)^{\bar{n}} - \frac{1}{\chi}\right] \right\}
\end{equation}
is a finite constant, which provides us the middle and bottom rows of Eq.~(\ref{equation: f asymptotic k}). Plugging $\bar{n}=2$ into Eq.~(\ref{equation: Cf general}) results with Eq.~(\ref{equation: Cf definition}).

\section{Leading behavior of $\hat{\psi}(u \to 0)$}
\label{appendix: derivation small u}

We assume that $\psi(\tau)$ possesses the asymptotic behavior
\begin{equation}
\label{equation: psi asymptotics tau}
	\lim_{\tau \rightarrow \infty} \psi(\tau)\tau^{1+\nu} = \tau_0^{\nu} ,
\end{equation}
where $\nu>0$ and $\tau_0>0$ are real numbers. The Laplace transform of $\psi(\tau)$ is defined as
\begin{equation}
\label{equation: laplace transform definition}
	\hat{\psi}(u) = \int_{0}^{\infty} \text{d}\tau \, \psi(\tau) e^{-u\tau} .
\end{equation}
As we show below, given a positive integer $\bar{n}$, Eqs.~(\ref{equation: psi asymptotics tau}-\ref{equation: laplace transform definition}) lead to
\begin{equation}
\label{equation: psi asymptotics u}
	\hat{\psi}(u) \simeq \sum_{n=0}^{\bar{n}-1} \frac{(-u)^n}{n!}\left<\tau^n\right> + \left\{
	\begin{aligned}
		& \Gamma(-\nu)(\tau_0u)^{\nu} & \bar{n}-1<\nu<\bar{n} \\
		& \frac{1}{\bar{n}!} (-1)^{\bar{n}+1} (\tau_0u)^{\bar{n}} \ln(C_{\bar{n}}\tau_0u) & \nu=\bar{n}
	\end{aligned}
	\right. ,
\end{equation}
where $C_{\bar{n}}\left[\psi(\tau)\right]$ is defined in the corresponding section. Using Eqs.~(\ref{equation: psi asymptotics tau}) and (\ref{equation: laplace transform definition}), we extract all of the converging moments out of the Laplace transform integral
\begin{equation}
\label{equation: laplace transform extracted}
	\hat{\psi}(u) = \sum_{n=0}^{\bar{n}-1} \frac{(-u)^n}{n!}\left<\tau^n\right> + \int_0^{\infty} \text{d}\tau \, \psi(\tau) \left[ e^{-u\tau} - \sum_{n=0}^{\bar{n}-1} \frac{(-u\tau)^n}{n!}\right] ,
\end{equation}
and consider each of the cases in Eq.~(\ref{equation: psi asymptotics u}) separately.

\subsection{A non-integer $\nu$}

Let us assume that $\tilde{n}-1<\nu<\tilde{n}$. In order to obtain the leading behavior of the second term of Eq.~(\ref{equation: laplace transform extracted}), we consider the limit
\begin{equation}
	l_0=\lim_{u \rightarrow 0} \frac{1}{u^{\nu}}\int_{0}^{\infty} \text{d}\tau \, \psi(\tau) \left[ e^{-u\tau} - \sum_{n=0}^{\bar{n}-1} \frac{(-u\tau)^n}{n!}\right] .
\end{equation}
Using L'Hospital's rule $\tilde{n}$ times yields
\begin{equation}
	l_0=\lim_{u \rightarrow 0} \frac{\Gamma(-\nu)u^{\bar{n}-\nu}}{\Gamma(-\nu+\bar{n})} \int_{0}^{\infty} \text{d}\tau \, \psi(\tau) \tau^{\bar{n}} e^{-u\tau} .
\end{equation}
Changing the integration variable to $\eta=u\tau$ produces
\begin{equation}
	l_0=\frac{\Gamma(-\nu)}{\Gamma(-\nu+\bar{n})} \lim_{u \rightarrow 0} \int_{0}^{\infty} \text{d}\eta \, \psi\left(\frac{\eta}{u}\right) \left(\frac{\eta}{u}\right)^{\nu+1} \eta^{\bar{n}-1-\nu} e^{-\eta} .
\end{equation}
We switch the order of limit and integration, which together with the asymptotic behavior (\ref{equation: psi asymptotics tau}) gives
\begin{equation}
	l_0=\frac{\tau_0^{\nu}\Gamma(-\nu)}{\Gamma(-\nu+\tilde{n})}\int_{0^+}^{\infty} \text{d}\eta \, \eta^{\tilde{n}-1-\nu} e^{-\eta} = \tau_0^{\nu}\Gamma(-\nu) .
\end{equation}
Which proves Eq.~(\ref{equation: psi asymptotics u})'s top row.

\subsection{An integer $\nu$}

Let us assume that $\nu=\bar{n}$, where $\bar{n}$ can be even or odd. In order to find the leading behavior of Eq.~(\ref{equation: laplace transform extracted}), we consider the following limit
\begin{equation}
	l_1=\lim_{u \rightarrow 0} \frac{1}{u^{\bar{n}}\ln(u)} \int_{0}^{\infty} \text{d}\tau \, \psi(\tau) \left[ e^{-u\tau} - \sum_{n=0}^{\bar{n}-1} \frac{(-u\tau)^n}{n!}\right] .
\end{equation}
Using L'Hospital's rule $\tilde{n}+1$, we get
\begin{equation}
	l_1=\lim_{u \rightarrow 0} (-1)^{\tilde{n}+1} \frac{u}{\tilde{n}!} \int_{0}^{\infty} \text{d}\tau \, \psi(\tau) \tau^{\tilde{n}+1} e^{-u\tau} .
\end{equation}
We change the integration variable to $\eta=u\tau$
\begin{equation}
	l_1=\lim_{u \rightarrow 0} \frac{(-1)^{\bar{n}+1}}{\bar{n}!} \int_{0}^{\infty} \text{d}\eta \, \psi\left(\frac{\eta}{u}\right) \left(\frac{\eta}{u}\right)^{\bar{n}+1} e^{-\eta} ,
\end{equation}
and switching the order of limit and integration, while applying the asymptotic behavior (\ref{equation: psi asymptotics tau})
\begin{equation}
	l_1=\frac{\tau_0^{\bar{n}}}{\bar{n}!} (-1)^{\bar{n}+1} \int_{0^{+}}^{\infty} \text{d}\eta \, e^{-\eta} = \frac{1}{\bar{n}!} (-1)^{\bar{n}+1} \tau_0^{\bar{n}} .
\end{equation}
To obtain the next order correction, we evaluate the following limit
\begin{equation}
	l_2=\lim_{u \rightarrow 0} \frac{1}{u^{\bar{n}}} \left\{ \int_{0}^{\infty} \psi(\tau) \left[ e^{-u\tau} - \sum_{n=0}^{\bar{n}-1} \frac{(-u\tau)^n}{n!}\right] \text{d}\tau - \frac{\tau_0^{\bar{n}}}{\bar{n}!} (-1)^{\bar{n}+1} u^{\bar{n}}\ln(u) \right\}.
\end{equation}
We use L'Hospital's rule $\bar{n}$ times to get
\begin{equation}
	l_2 = \frac{(-1)^{\bar{n}}}{\bar{n}!} \lim_{u \to 0} \left\{\vphantom{\int_0^{\infty}} \int_0^{\infty} \text{d}\tau \, \psi(\tau) \tau^{\bar{n}} e^{-u\tau} + \tau_0^{\bar{n}} \left[ \ln(u) + H_{\bar{n}} \vphantom{\frac{1}{1}} \right] \right\},
\end{equation}
where $H_{\bar{n}}=\sum_{n=1}^{\bar{n}} \frac{1}{n}$ is the $\bar{n}$th harmonic number. Splitting the integral at $\tau=\tau_0$ yields
\begin{equation}
\label{equation: psi integer limit 1}
	l_2 = \frac{(-1)^{\bar{n}}}{\bar{n}!} \lim_{u \to 0} \left\{ \int_0^{\tau_0} \text{d}\tau \, \psi(\tau) \tau^{\bar{n}} e^{-u\tau} + \int_{\tau_0}^{\infty} \text{d}\tau \, \psi(\tau) \tau^{\bar{n}} e^{-u\tau} + \tau_0^{\bar{n}} \left[ \ln(u) + H_{\bar{n}} \vphantom{\frac{1}{1}} \right] \right\} .
\end{equation}
We now add and subtract the term $\tau_0^{\bar{n}}/\tau^{\bar{n}+1}$ from $\psi(\tau)$ in the second integral in Eq.~(\ref{equation: psi integer limit 1})
\begin{equation}
\label{equation: psi integer limit 2}
	l_2=\frac{(-1)^{\bar{n}}}{\bar{n}!} \lim_{u \rightarrow 0} \left\{ \int_{0}^{\tau_0} \text{d}\tau \, \psi(\tau) \tau^{\bar{n}} e^{-u\tau} + \int_{\tau_0}^{\infty} \text{d}\tau \left[ \psi(\tau) - \frac{\tau_0^{\bar{n}}}{\tau^{\bar{n}+1}}\right] \tau^{\bar{n}} e^{-u\tau} + \int_{\tau_0}^{\infty} \text{d}\tau \, \frac{\tau_0^{\bar{n}}}{\tau} e^{-u\tau} + \tau_0^{\bar{n}} \left[ \ln(u) + H_{\bar{n}} \vphantom{\frac{1}{1}} \right] \right\} \nonumber,
\end{equation}
Note that due to the asymptotics (\ref{equation: psi asymptotics tau}), the middle integral is finite when $u \to 0$. Eq.~(\ref{equation: psi integer limit 2})'s right integral can be computed explicitly, after which the limit can be evaluated. Finally, we find
\begin{equation}
	l_2=\frac{1}{\bar{n}!} (-1)^{\bar{n}+1} \tau_0^{\bar{n}} \ln(C_{\bar{n}}\tau_0) ,
\end{equation}
where
\begin{equation}
\label{equation: Cpsi general}
	C_{\bar{n}}\left[\psi(\tau)\right] = \exp \left\{ \gamma - H_{\tilde{n}} - \int_{0}^{\tau_0} \text{d}\tau \, \psi(\tau) \left(\frac{\tau_0}{\tau}\right)^{\bar{n}} - \int_{\tau_0}^{\infty} \text{d}\tau \left[ \psi(\tau)\left(\frac{\tau_0}{\tau}\right)^{\bar{n}} - \frac{1}{\tau} \right] \right\}
\end{equation}
is a finite constant, which concludes the bottom row of Eq.~(\ref{equation: psi asymptotics u}). Plugging $\bar{n}=2$ into Eq.~(\ref{equation: Cpsi general}) results with Eq.~(\ref{equation: Cpsi definition}).

\section{Additional steps in the derivation of Lambert scaling for the L\'evy walk model}
\label{appendix: additional details for lambert levy}

Here we portray the additional steps which were omitted in section \ref{section: labert scaling for levy walk}. Starting from Eq.~(\ref{equation: small u behavior}), we expand $\hat{\psi}(u-i\boldsymbol{k}\cdot\boldsymbol{v})$ for small arguments
\begin{equation}
	\hat{\psi}\left(u-i\boldsymbol{k}\cdot\boldsymbol{v}\right) \simeq 1 - \left<\tau\right>\left(u-i\boldsymbol{k}\cdot\boldsymbol{v}\right) - \frac{1}{2}\tau_0^2\left(u-i\boldsymbol{k}\cdot\boldsymbol{v}\right)^2 \ln\left[C_{\psi}\tau_0\left(u-i\boldsymbol{k}\cdot\boldsymbol{v}\right)\right] ,
\end{equation}
and apply this expansion to the Montroll-Weiss Eq.~(\ref{equation: montroll weiss})
\begin{equation}
\label{equation: montroll weiss expansion 1}
	\Pi_d\left(\boldsymbol{k}, u\right) \simeq \left\{ 1 + \left< \frac{\tau_0^2}{2\langle\tau\rangle}  \left(u-i\boldsymbol{k}\cdot\boldsymbol{v}\right) \ln\left[C_{\psi}\tau_0\left(u-i\boldsymbol{k}\cdot\boldsymbol{v}\right) \vphantom{\frac{\tau_0^2}{2\langle\tau\rangle}} \right] \right> \right\} \left\{ u + \left< \frac{\tau_0^2}{2\langle\tau\rangle}  \left(u-i\boldsymbol{k}\cdot\boldsymbol{v}\right)^2 \ln\left[C_{\psi}\tau_0\left(u-i\boldsymbol{k}\cdot\boldsymbol{v}\right) \vphantom{\frac{\tau_0^2}{2\langle\tau\rangle}} \right] \right> \right\}^{-1} .
\end{equation}
We use the identity
\begin{equation}
	\ln\left(a\pm ib\right) = \frac{1}{2}\ln\left(a^2+b^2\right) \pm i\tan^{-1}\left(\frac{b}{a}\right) ,
\end{equation}
and the assumed symmetry of $F(\boldsymbol{v})$ in order to simplify Eq.~(\ref{equation: montroll weiss expansion 1})
\begin{align}
\label{equation: montroll weiss expansion 2}
	&\Pi_d\left(\boldsymbol{k}, u\right) \simeq \\
	&\left\{ 1 + \frac{u\tau_0^2}{4\langle\tau\rangle} \left< \ln\left[C_{\psi}^2\tau_0^2\left[u^2+(\boldsymbol{k}\cdot\boldsymbol{v})^2\right] \vphantom{\frac{\tau_0^2}{2\langle\tau\rangle}} \right] \right> - \frac{\tau_0^2}{2\langle\tau\rangle} \left< (\boldsymbol{k}\cdot\boldsymbol{v}) \tan^{-1}\left[ \frac{\boldsymbol{k}\cdot\boldsymbol{v}}{u} \vphantom{\frac{\tau_0^2}{2\langle\tau\rangle}} \right] \right> \right\} \times \nonumber \\
	&\left\{ u + \frac{u^2\tau_0^2}{4\langle\tau\rangle} \left< \ln\left[C_{\psi}^2\tau_0^2\left[u^2+(\boldsymbol{k}\cdot\boldsymbol{v})^2\right] \vphantom{\frac{\tau_0^2}{2\langle\tau\rangle}} \right] \right> - \frac{\tau_0^2}{4\langle\tau\rangle} \left< (\boldsymbol{k}\cdot\boldsymbol{v})^2 \ln\left[C_{\psi}^2\tau_0^2\left[u^2+(\boldsymbol{k}\cdot\boldsymbol{v})^2\right] \vphantom{\frac{\tau_0^2}{2\langle\tau\rangle}} \right] \right> + \frac{u\tau_0^2}{\langle\tau\rangle} \left< (\boldsymbol{k}\cdot\boldsymbol{v}) \tan^{-1}\left[ \frac{\boldsymbol{k}\cdot\boldsymbol{v}}{u} \vphantom{\frac{\tau_0^2}{2\langle\tau\rangle}} \right] \right> \right\}^{-1} . \nonumber
\end{align}
Using $\ln(1+\epsilon^2)\simeq\epsilon^2$ and $\tan^{-1}(1/\epsilon)\simeq(\pi/2)\text{sign}(\epsilon)-\epsilon$ for $\epsilon\to0$, we discard irrelevant terms with respect to $\epsilon\sim u/kv$, so we have
\begin{equation}
\label{equation: montroll weiss expansion 3}
	\Pi_d\left(\boldsymbol{k}, u\right) \simeq \left\{ u - \frac{\tau_0^2}{4\langle\tau\rangle} \left< (\boldsymbol{k}\cdot\boldsymbol{v})^2 \ln\left[C_{\psi}^2\tau_0^2(\boldsymbol{k}\cdot\boldsymbol{v})^2 \vphantom{\frac{\tau_0^2}{2\langle\tau\rangle}} \right] \right> \right\}^{-1} .
\end{equation}
One may argue that this expansion breaks down when $v=0$ or alternatively when $\boldsymbol{k} \cdot \boldsymbol{v}=0$, but actually there is no problem. The former case is ruled out since our physical models have a positive constant for the speed $v=V=1$, and as such we demand that $F_d(v\ne1)=0$. The latter case is ruled out since $F_d(\boldsymbol{v})$ covers all velocity directions of the $d$-dimensional space, by construction. Therefore, it will always contain a part parallel to $\boldsymbol{k}$, regardless of $\boldsymbol{k}$'s direction. Returning to the time domain, we obtain
\begin{equation}
	\tilde{P}_d(\boldsymbol{k},t) \simeq \exp\left\{ \frac{t\tau_0^2}{4\langle\tau\rangle} \left< (\boldsymbol{k}\cdot\boldsymbol{v})^2 \ln\left[C_{\psi}^2\tau_0^2(\boldsymbol{k}\cdot\boldsymbol{v})^2 \vphantom{\frac{\tau_0^2}{2\langle\tau\rangle}} \right] \right> \right\} .
\end{equation}
Substituting $\boldsymbol{\kappa}=\boldsymbol{k} \sqrt{\tau_0^2\langle v^2\rangle t\Omega_d(t)/4d\left<\tau\right>}$, where $\Omega_d(t)$ is a scaling function, leads to
\begin{equation}
\label{equation: fourier expansion 1}
	\tilde{P}_d(\boldsymbol{\kappa},t) \simeq \left[\frac{4d\left<\tau\right>}{\tau_0^2\langle v^2 \rangle t\Omega(t)}\right]^{d/2} \exp\left\{ \frac{1}{\Omega(t)} \left< \frac{(\boldsymbol{\kappa}\cdot\boldsymbol{v})^2}{\langle v^2 \rangle/d} \ln\left[\frac{4dC_{\psi}^2\langle\tau\rangle}{t \Omega(t)} \frac{(\boldsymbol{\kappa}\cdot\boldsymbol{v})^2}{\langle v^2 \rangle} \right] \right> \right\} .
\end{equation}
We determine the slowly varying scaling function $\Omega_d(t)$ by demanding that $\ln[t\Omega_d(t)/(4dC_{\psi}^2\langle\tau\rangle)]=\Omega_d(t)$, obtaining Eq.~(\ref{equation: omega and xi definition for t}). Thus, Eq.~(\ref{equation: fourier expansion 1}) becomes
\begin{align}
\label{equation: fourier expansion 2}
	\tilde{P}_d(\boldsymbol{\kappa},t) &\simeq \left[\frac{2}{\xi_d(t)}\right]^d \exp\left[ - \left< \frac{(\boldsymbol{\kappa}\cdot\boldsymbol{v})^2}{\langle v^2 \rangle/d} \right> \right] \exp\left\{ \frac{1}{\Omega_d(t)} \left< \frac{(\boldsymbol{\kappa}\cdot\boldsymbol{v})^2}{\langle v^2 \rangle/d} \ln\left[ \frac{(\boldsymbol{\kappa}\cdot\boldsymbol{v})^2}{\langle v^2 \rangle} \right] \right> \right\} \nonumber \\
	& \simeq \left[\frac{2}{\xi_d(t)}\right]^d \exp\left[ - \left< \frac{(\boldsymbol{\kappa}\cdot\boldsymbol{v})^2}{\langle v^2 \rangle/d} \right> \right] \left\{ 1 + \frac{1}{\Omega(t)} \left< \frac{(\boldsymbol{\kappa}\cdot\boldsymbol{v})^2}{\langle v^2 \rangle/d} \ln\left[ \frac{(\boldsymbol{\kappa}\cdot\boldsymbol{v})^2}{\langle v^2 \rangle} \right] \right> \right\} ,
\end{align}
where we expanded the exponential term in the second row due to the assumption of large $t$, which leads to large $\Omega_d(t)$. Equations~(\ref{equation: fourier transform of the pipe distribution}) and (\ref{equation: fourier transform of the gas distribution}) then follow from Eqs.~(\ref{equation: velocity distibution for the pipe}) and (\ref{equation: velocity distibution for the gas}), respectively, when combined with Eq.~(\ref{equation: fourier expansion 2}).

\section{Derivation of the integration boundaries}
\label{appendix: integration boundaries calculation}

\subsection{Lorentz gas model}

We denote as $(x_0,y_0)$ the starting point on the $(0,0)$ scatterer from which we assume the particle has originated. The pair $\{b,\beta\}$ and the trio $\{x_0,y_0,\beta\}$ are related by a simple transformation. To obtain it, we define the two vectors
\begin{equation}
	\boldsymbol{R} = x_0\hat{x} + y_0\hat{y} , \quad \boldsymbol{B} = -b\sin(\beta)\hat{x} + b\cos(\beta)\hat{y} ,
\end{equation}
where $\boldsymbol{B}$ can be seen in Fig.~\ref{ParametersDefinitionsCDF} (the red arrow). Solving the equation $\boldsymbol{B}\cdot(\boldsymbol{R}-\boldsymbol{B})=0$ and $x_0(0,0,R)=R$, we get the following expressions
\begin{align}
	x_0(b,\beta,R) &= \sqrt{R^2-b^2}\cos(\beta) - b\sin(\beta) , \nonumber \\
	y_0(b,\beta,R) &= \sqrt{R^2-b^2}\sin(\beta) + b\cos(\beta) .
\end{align}
By solving
\begin{equation}
	\left[x_0(b,\beta,R)+\tau_{n,m}^*(b,\beta,R)\cos(\beta)-n\right]^2 + \left[y_0(b,\beta,R)+\tau_{n,m}^*(b,\beta,R)\sin(\beta)-m\right]^2 = R^2 ,
\end{equation}
together with $\tau_{1,0}^*(0,0,R) = 1-2R$, we obtain Eqs.~(\ref{equation: tau star definition}) and (\ref{equation: tau star discriminant}). In order to extract the integration boundaries (IBs) of $\beta$ and $b$ out of the inequalities Eqs.~(\ref{equation: cross NN inequality}-\ref{equation: cross DN inequalities}), we notice that there are two classes of trajectories which reach the $(n,1)$ scatterer, where $n\ge0$. The first class' IBs are dictated by the origin and target scatterers. The second class' IBs are governed by the $(1,0)$ and $(n-1,1)$ scatterers [this class does not exist for the $(0,1)$ circle]. We split the $\beta$ domain into two parts marked (I) and (II), each corresponds to a class of trajectories, and denote the separator angle between them with $\beta^{\rm sep}_{n,1}(R)$. Each of $\beta$'s subdomains is associated with a different expression for $b$'s IBs, which we denote (i) and (ii). However, when $\beta=\beta^{\rm sep}_{n,1}(R)$ the two expressions coincide. Therefore, $\beta^{\rm sep}_{n,1}(R)$ can be found by comparing the lower (or upper) IB of (i) to that of (ii). Finding the top/bottom IBs of $\beta$ relies on a similar idea, namely using the expressions for $b$'s IBs. When $\beta=\beta^{\rm min}_{n,1}(R)$, the associated $b$ subdomain (i) shrinks to zero. Thus, $\beta^{\rm min}_{n,1}(R)$ can be found by setting $b^{\rm min}_{n,1}(\beta,R)=b^{\rm max}_{n,1}(\beta,R)$ on the (i) expression. Identically, $\beta^{\rm max}_{n,1}(R)$ is found by setting $b^{\rm min}_{n,1}(\beta,R)=b^{\rm max}_{n,1}(\beta,R)$ on the (ii) expression. Starting with $n=0$, the upper IB of $\beta$ is set to $\pi/2$, which is possible due to symmetry. This is done in order to balance out the excess of distant neighbors groups over nearest and next to nearest neighbors groups across the lattice, which is demonstrated in Fig.~\ref{LorentzSymmetry}. Writing Eq.~(\ref{equation: cross NN inequality}) in its explicit form, we get
\begin{equation}
\label{equation: cross NN inequality explicit}
	\text{(i) } \cos(\beta)-R \le b \le \cos(\beta)+R .
\end{equation}
Since $\beta\le\pi/2$ and $|b|\le R$, the upper IB in Eq.~(\ref{equation: cross NN inequality explicit}) is set to $R$. For $\beta$'s lower IB, we equate $b$'s IBs, $\cos(\beta)-R=R$, getting $\cos^{-1}(2R)$. Thus, we obtain for $n=0$
\begin{equation}
	\frac{\pi}{2} - \sin^{-1}(2R) \le \beta \le \frac{\pi}{2} , \quad \cos(\beta)-R \le b \le R .
\end{equation}
We continue with $n=1$. Again, due to symmetry we can truncate $\beta$'s upper IB to $\pi/4$. Equation~(\ref{equation: cross NNN inequalities}) in its explicit form combined with $|b|\le R$ then yields
\begin{equation}
\label{equation: cross NNN inequalities explicit}
	\text{(i) } \cos(\beta)-\sin(\beta)-R \le b \le R , \quad \text{(ii) } R-\sin(\beta) \le b \le \cos(\beta)-R .
\end{equation}
Here, the $\beta$ domain is sectioned into two parts, as said before: (I) for which the (ii) inequality in Eq.~(\ref{equation: cross NNN inequalities explicit}) is trivially fulfilled, and (II) in which it needs to be upheld. Equating the IBs of (i) in Eq.~(\ref{equation: cross NNN inequalities explicit}) yields the lower IB of $\beta$. We obtain
\begin{equation}
	\frac{\pi}{4} - \sin^{-1}\left(\sqrt{2}R\right) \le \beta \le \frac{\pi}{4} .
\end{equation}
The separator angle between (I) and (II) can be found by comparing the lower (or upper) $b$ IB of (i) to that of (ii), yielding $\beta=\cos^{-1}(2R)$. Therefore we find for $n=1$
\begin{equation}
	\text{(I) } \frac{\pi}{4} - \sin^{-1}\left(\sqrt{2}R\right) \le \beta \le \cos^{-1}(2R) , \quad \text{(i) } \cos(\beta)-\sin(\beta)-R \le b \le R ,
\end{equation}
and
\begin{equation}
	\text{(II) } \cos^{-1}(2R) \le \beta \le \frac{\pi}{4} , \quad \text{(ii) } R-\sin(\beta) \le b \le \cos(\beta)-R .
\end{equation}
Finally, Eq.~(\ref{equation: cross DN inequalities}) in its explicit form supply
\begin{equation}
\label{equation: cross DN inequalities explicit}
	\text{(i) } \cos(\beta)-n\sin(\beta)-R \le b \le R , \quad \text{(ii) } R-\sin(\beta) \le b \le \cos(\beta)-(n-1)\sin(\beta)-R ,
\end{equation}
where we already implemented $|b|\le R$ to (i)'s top IB. This time, there is no need to set the upper IB of $\beta$ manually. For the bottom IB of $\beta$'s subdomain (I), one equates Eq.~(\ref{equation: cross DN inequalities explicit})'s (i) IBs, namely $\cos(\beta)-n\sin(\beta) = 2R$. By dividing this equality with $\sqrt{n^2+1}$ and using basic trigonometry, we have a general form of the lower IB of $\beta$, Eq.~(\ref{equation: cross lower beta boundary}). The same is done for the upper IB of $\beta$'s subdomain (II), where (ii) of Eq.~(\ref{equation: cross DN inequalities explicit}) is used, yielding $\cos(\beta)-(n-2)\sin(\beta)=2R$, which results with Eq.~(\ref{equation: cross upper beta boundary}). The separator angle can be found by equating the top/bottom IBs of (i) to (ii), producing $\cos(\beta)-(n-1)\sin(\beta)=2R$ and Eq.~(\ref{equation: cross beta separator}). We finish this subsection with the case of $R=0.3$, namely four open corridors. As said in section \ref{section: cdf calculation}, the scatterer $(2,1)$ is now shared by the horizontal and diagonal directions, thus we adjust its upper $\beta$ IB. When $\beta$ achieves its maximal value for the horizontal direction stripe, $b$'s lower IB must be equal to $-R$. Thus we set $b^{\rm min}_{2,1}(\beta,R)=-R$ in Eq.~(\ref{equation: cross b boundaries}), and obtain Eq.~(\ref{equation: cross upper beta boundary correction}). Similarly to the horizontal case, there are two classes of trajectories to reach the $(m+1,m)$ scatterer. The first class' IBs are governed by the origin and target scatterers, and the second class' IBs are dictated by the $(1,1)$ and $(m,m-1)$ circles [this class does not exist for the $(2,1)$ target scatterer]. Again, we split $\beta$'s domain into two subdomains (I) and (II), however this time they switch places, i.e. the (I) part is of higher $\beta$ values than the (II) part. Equation~(\ref{equation: flag NN inequality}) in its explicit form reads
\begin{equation}
\label{equation: flag NN inequality explicit}
	\text{(i) } -R \le b \le \cos(\beta)-2\sin(\beta)+R ,
\end{equation}
where we set the lower $b$ IB to $-R$ since $\beta\ge\sin^{-1}(2R)$ and $|b|\le R$. For the upper IB of $\beta$ we equate the IBs of $b$ in Eq.~(\ref{equation: flag NN inequality explicit}) to each other, namely $-R=\cos(\beta)-2\sin(\beta)+R$. Thus we obtain for $m=1$
\begin{equation}
	\sin^{-1}(2R) \le \beta \le \sin^{-1}\left(\frac{4R}{5}+\frac{1}{5}\sqrt{5-4R^2}\right) .
\end{equation}
For $m>1$, we write Eq.~(\ref{equation: flag DN inequalities}) in its explicit form
\begin{equation}
\label{equation: flag DN inequalities explicit}
	\text{(i) } -R \le b \le m\cos(\beta)-(m+1)\sin(\beta)+R , \quad \text{(ii) } (m-1)\cos(\beta)-m\sin(\beta)+R \le b \le \cos(\beta)-\sin(\beta)-R ,
\end{equation}
where we set the bottom IB of (i) to $-R$, as before. The upper $\beta$ IB is found by equating (i)'s top IB to $-R$ in Eq.~(\ref{equation: flag DN inequalities explicit}), and the bottom $\beta$ IB is found by equating (ii)'s top and bottom IBs to each other. The separator angle is found in the same manner as for the horizontal case. Dividing the resulted expressions with $\sqrt{m^2+(m+1)^2}$ and using basic trigonometry, we obtain Eqs.~(\ref{equation: flag upper beta boundary}-\ref{equation: flag beta separator}).

\subsection{Stadium channel model}

We denote as $(x_0,y_0)$ the starting point on the $(0,0)$ stadium from which we assume the particle has originated. The pair $\{a,\alpha\}$ and the trio $\{x_0,y_0,\alpha\}$ are related by a simple transformation. To obtain it, we define the two vectors
\begin{equation}
	\boldsymbol{R} = x_0\hat{x} + y_0\hat{y} , \quad \boldsymbol{A} = -a\sin(\alpha)\hat{x} + a\cos(\alpha)\hat{y} ,
\end{equation}
where $\boldsymbol{A}$ can be seen in Fig.~\ref{ParametersDefinitionsCDF} (the red arrow). Solving the equation $\boldsymbol{A}\cdot(\boldsymbol{R}-\boldsymbol{A})=0$ and $y_0(0,0)=-1$, we get the following expressions
\begin{align}
	x_0(a,\alpha) &= \sqrt{1-a^2}\cos(\alpha) - a\sin(\alpha) , \nonumber \\
	y_0(a,\alpha) &= \sqrt{1-a^2}\sin(\alpha) + a\cos(\alpha) .
\end{align}
By solving
\begin{equation}
	\left[x_0(a,\alpha)+\tau_{2n,m}^*(a,\alpha)\cos(\alpha)-2n\right]^2 + \left[y_0(a,\alpha)+\tau_{2n,m}^*(a,\alpha)\sin(\alpha)-m\right]^2 = 1 ,
\end{equation}
together with $\tau_{0,0}^*(0,0) = 2$, we obtain Eqs.~(\ref{equation: tau star definition pipe}) and (\ref{equation: y component origin target points}). Simplifying Eq.~(\ref{equation: pipe NN inequalities}) and using $-\pi/2\le\alpha\le\pi/2$, we get Eq.~(\ref{equation: pipe origin as target boundaries}). In order to extract $\alpha$'s lower IB out of the inequalities Eq.~(\ref{equation: pipe NNN inequalities}), we notice that as with the Lorentz gas model, when $\alpha$ hits its bottom IB, the $a$ domain vanishes. Simplifying Eq.~(\ref{equation: pipe NNN inequalities}), we have
\begin{equation}
	D\cos(\alpha)-\sin(\alpha) \le a \le \sin(\alpha) ,
\end{equation}
and thus we have $D\cos(\alpha)-\sin(\alpha)=\sin(\alpha)$ for $\alpha$'s lower IB, which solves to $\alpha_{0,D}^{\rm min}=\tan^{-1}(D/2)$. Finally, we notice that there are two classes of trajectories which reach the target stadiums of $n>0$, as with the Lorentz gas case. The first class' IBs are dictated by the origin and target semicircles. The second class' IBs are governed by the origin and $(2n-2,D)$ stadium walls [this class does not exist for the trajectories ending with the origin or $(0,D)$ semicircles]. We split the $\alpha$ domain into two parts marked (I) and (II), each corresponds to a class of trajectories, and denote the separator angle between them with $\alpha^{\rm sep}_{2n,D}$. Each of $\alpha$'s subdomains is associated with a different expression for $a$'s IBs, which we denote (i) and (ii). However, when $\alpha=\alpha^{\rm sep}_{2n,D}$ the two expressions coincide. Therefore, $\alpha^{\rm sep}_{2n,D}$ can be found by comparing the lower (or upper) IB of (i) to that of (ii). Finding the top/bottom IBs of $\alpha$ relies on a similar idea, namely using the expressions for $a$'s IBs. When $\alpha=\alpha^{\rm min}_{2n,D}$, the associated $a$ subdomain (i) shrinks to zero. Thus, $\alpha^{\rm min}_{2n,D}$ can be found by setting $a^{\rm min}_{2n,D}(\alpha)=a^{\rm max}_{2n,D}(\alpha)$ on the (i) expression. Identically, $\alpha^{\rm max}_{2n,D}$ is found by setting $a^{\rm min}_{2n,D}(\alpha)=a^{\rm max}_{2n,D}(\alpha)$ on the (ii) expression. Simplifying Eq.~(\ref{equation: pipe DN inequalities})'s top row, we get
\begin{equation}
\label{equation: pipe DN inequalities simplified top}
	\text{(i) } -\sin(\alpha) \le a \le \sin(\alpha) , \quad D\cos(\alpha)-(2n+1)\sin(\alpha) \le a \le D\cos(\alpha)-(2n-1)\sin(\alpha) .
\end{equation}
This class of trajectories has its upper/lower $a$ IB dominated by the origin/target stadium, and thus the top/bottom IB for this class is taken from Eq.~(\ref{equation: pipe DN inequalities simplified top})'s first/second inequality, such that
\begin{equation}
	\text{(i) } D\cos(\alpha)-(2n+1)\sin(\alpha) \le a \le \sin(\alpha) .
\end{equation}
For the inequalities of Eq.~(\ref{equation: pipe DN inequalities})'s bottom row, we get after simplification
\begin{equation}
	\text{(ii) } -\sin(\alpha) \le a \le \sin(\alpha) , \quad \left| a-D\cos(\alpha)+(2n-2)\sin(\alpha) \right| \ge \sin(\alpha) .
\end{equation}
This class of trajectories has its lower/upper $a$ IB dominated by the origin/$(2n-2,D)$ stadium, and thus the bottom/top IB for this class is taken from Eq.~(\ref{equation: pipe DN inequalities simplified top})'s first/second inequality, such that
\begin{equation}
	\text{(ii) } -\sin(\alpha) \le a \le D\cos(\alpha)-(2n-1)\sin(\alpha) .
\end{equation}
Using the IBs of $a$ in (i) and (ii) to extract the IBs of $\alpha$ in the way described above, we obtain Eqs.~(\ref{equation: pipe lower alpha boundary}-\ref{equation: pipe alpha separator}).

\section{Calculations of $\tau_0$, $\langle\tau\rangle$, and $C_{\psi}$}
\label{appendix: constants calculation}

\subsection{Lorentz gas model}

Using Eq.~(\ref{equation: psi as sum of integrals}) together with the integration boundaries Eqs.~(\ref{equation: cross lower beta boundary}-\ref{equation: cross beta separator}), one can calculate $\psi(\tau)$ using a computational program like Mathematica and extract the constants $\tau_0$, $\langle\tau\rangle$, and $C_{\psi}$ out of it. However, we found that an analytical expression for $\tau_0$ can be calculated. It follows from its definition Eq.~(\ref{equation: fat tail}) that
\begin{equation}
\label{equation: fat tail integrated}
	\tau_0^2 = \lim_{T\to\infty} \frac{1}{T} \int_{0}^{T} \text{d}\tau \, \tau^3 \, \psi(\tau) .
\end{equation}
Plugging Eq.~(\ref{equation: psi as sum of integrals}) with two infinite corridors (i.e. $q=0$) into Eq.~(\ref{equation: fat tail integrated}), we get
\begin{equation}
\label{equation: fat tail summed zero q}
	\tau_0^2 = 8\lim_{T\to\infty} \frac{1}{T} \sum_{n=0}^{\infty} \int_{\beta_{n,1}^{\rm min}(R)}^{\beta_{n,1}^{\rm max}(R)} \frac{\text{d}\beta}{2\pi} \int_{b_{n,1}^{\rm min}(\beta,R)}^{b_{n,1}^{\rm max}(\beta,R)} \frac{\text{d}b}{2R} \tau^{*3}_{n,1}(b,\beta,R)\text{H}\left[ T - \tau^*_{n,1}(b,\beta,R) \vphantom{\frac{1}{2}} \right] ,
\end{equation}
where $\text{H}(\cdots)$ is the Heaviside step function, and the factor of $8$ arise from symmetry, see Fig.~(\ref{LorentzSymmetry}). Since $\tau_{n,m}^*(b,\beta,R)$ is the traveling distance to the $(n,m)$ scatterer, it obeys $\tau_{n,1}^*(b,\beta,R)\simeq\sqrt{n^2+(1-2R)^2}\simeq n$ when $n$ is large, thus the Heaviside function truncates the sum in Eq.~(\ref{equation: fat tail summed zero q}) at $n=T$, and we have
\begin{equation}
\label{equation: fat tail summed zero q truncated}
	\tau_0^2 = 8\lim_{T\to\infty} \frac{1}{T} \sum_{n=0}^{T} \int_{\beta_{n,1}^{\rm min}(R)}^{\beta_{n,1}^{\rm max}(R)} \frac{\text{d}\beta}{2\pi} \int_{b_{n,1}^{\rm min}(\beta,R)}^{b_{n,1}^{\rm max}(\beta,R)} \frac{\text{d}b}{2R} \tau^{*3}_{n,1}(b,\beta,R) .
\end{equation}
Equation~(\ref{equation: fat tail integrated}) suggests that the integral over $\tau^3\psi(\tau)$ grows linearly with $T$, and as such we expect that the sum in Eq.~(\ref{equation: fat tail summed zero q truncated}) will behave similarly with $T$. We therefore write
\begin{equation}
	\tau_0^2 = 8\lim_{T\to\infty} \int_{\beta_{T,1}^{\rm min}(R)}^{\beta_{T,1}^{\rm max}(R)} \frac{\text{d}\beta}{2\pi} \int_{b_{T,1}^{\rm min}(\beta,R)}^{b_{T,1}^{\rm max}(\beta,R)} \frac{\text{d}b}{2R} \left[\sqrt{T^2+(1-2R)^2}\right]^3 .
\end{equation}
Now the integrals can be easily performed. After evaluating the limit we get a closed expression for $\tau_0$, Eq.~(\ref{equation: fat tail tau zero}). The remaining constants $\langle\tau\rangle$ and $C_{\psi}$ are defined via integrals over $\psi(\tau)$ rather than by a limit operation, and as such they cannot be obtained using end terms as we just did. Even though, one is not required to calculate $\psi(\tau)$, but can use a simpler tactic. To calculate the mean time between collisions, we use its definition and plug inside Eq.~(\ref{equation: psi as sum of integrals})
\begin{equation}
	\langle\tau\rangle = 8\sum_{n=0}^{\infty} \int_{\beta_{n,1}^{\rm min}(R)}^{\beta_{n,1}^{\rm max}(R)} \frac{\text{d}\beta}{2\pi} \int_{b_{n,1}^{\rm min}(\beta,R)}^{b_{n,1}^{\rm max}(\beta,R)} \frac{\text{d}b}{2R}\tau^*_{n,1}(b,\beta,R) ,
\end{equation}
To achieve a designated precision, one can simply truncate the sum at a large enough $T$. For $T=500$ we obtain $\langle\tau\rangle\approx 0.62155$. For $C_{\psi}$ we find the following formula out of Eq.~(\ref{equation: Cpsi definition})
\begin{equation}
	C_{\psi} = \lim_{T\to\infty} \exp\left[ \gamma - \frac{3}{2} + \ln\left( \frac{T}{\tau_0} \right) - \int_0^T\text{d}\tau \, \psi(\tau) \left( \frac{\tau}{\tau_0} \right)^2 \right] .
\end{equation}
Plugging Eq.~(\ref{equation: psi as sum of integrals}) into the above expression and allowing the Heaviside function to truncate the sum at $n=T$ yields
\begin{equation}
	C_{\psi} = \lim_{T\to\infty} \exp\left[ \gamma - \frac{3}{2} + \ln\left( \frac{T}{\tau_0} \right) - 8\sum_{n=0}^T \int_{\beta_{n,1}^{\rm min}(R)}^{\beta_{n,1}^{\rm max}(R)} \frac{\text{d}\beta}{2\pi} \int_{b_{n,1}^{\rm min}(\beta,R)}^{b_{n,1}^{\rm max}(\beta,R)} \frac{\text{d}b}{2R} \frac{\tau^{*2}_{n,1}(b,\beta,R)}{\tau_0^2} \right] .
\end{equation}
This equation converges rather slowly due to the logarithmic term, hence we need to accelerate its convergence rate. To do that, we use the identity
\begin{equation}
	\lim_{T\to\infty}\left[\ln(T)-\sum_{n=1}^{T}\frac{1}{n}\right] = -\gamma ,
\end{equation}
and write, taking the $n=0$ summand out of the sum
\begin{align}
	C_{\psi} &= \lim_{T\to\infty} \exp\left\{ - \frac{3}{2} - \ln\left( \tau_0 \right) \vphantom{\sum_{n=0}^T \int_{\beta_{n,1}^{\rm min}(R)}^{\beta_{n,1}^{\rm max}(R)}} - 8\int_{\beta_{0,1}^{\rm min}(R)}^{\beta_{0,1}^{\rm max}(R)} \frac{\text{d}\beta}{2\pi} \int_{b_{0,1}^{\rm min}(\beta,R)}^{b_{0,1}^{\rm max}(\beta,R)} \frac{\text{d}b}{2R} \frac{\tau^{*2}_{0,1}(b,\beta,R)}{\tau_0^2} \right. \nonumber \\
	&-\left. \sum_{n=1}^T \left[8 \int_{\beta_{n,1}^{\rm min}(R)}^{\beta_{n,1}^{\rm max}(R)} \frac{\text{d}\beta}{2\pi} \int_{b_{n,1}^{\rm min}(\beta,R)}^{b_{n,1}^{\rm max}(\beta,R)} \frac{\text{d}b}{2R} \frac{\tau^{*2}_{n,1}(b,\beta,R)}{\tau_0^2} - \frac{1}{n} \right] \right\} .
\end{align}
We approximate $\tau_{n,1}^*(b,\beta,R)$ for large $n$ as before and see, after performing the integrals, that the summands behave for $n\gg 1$ as
\begin{equation}
	s_n \simeq 8\int_{\beta_{n,1}^{\rm min}(R)}^{\beta_{n,1}^{\rm max}(R)} \frac{\text{d}\beta}{2\pi} \int_{b_{n,1}^{\rm min}(\beta,R)}^{b_{n,1}^{\rm max}(\beta,R)} \frac{\text{d}b}{2R} \frac{n^2+(1-2R)^2}{\tau_0^2} -\frac{1}{n} \simeq \frac{B_1}{n^2} + \frac{B_2}{n^3} + \frac{B_3}{n^4} ,
\end{equation}
where $B_1,B_2,B_3$ are some $R$-dependent coefficients. This behavior suggests that the partial sum $S_T$ goes like
\begin{align}
\label{equation: richardson justification}
	S_T = \sum_{n=1}^T s_n \simeq l + \frac{\tilde{B}_1}{T} + \frac{\tilde{B}_2}{T^2} + \frac{\tilde{B}_3}{T^3} ,
\end{align}
for large $T$, where $l$ is the desired limit. This implies that the Richardson extrapolation method can be used, see Eq.~(8.1.16) in Ref.~\cite{Bender}. For $T=40$, we get $C_{\psi}\approx 4.4802\times 10^{-4}$ by summing terms up to $n=T$ and extrapolating over the $S_{37},S_{38},S_{39},S_{40}$ partial sums.

Extracting the needed constants for four open horizons, namely $q>0$, is done in a similar fashion, but this time we have $1/\sqrt{20} \le R < 1/\sqrt{8}$. Plugging Eq.~(\ref{equation: psi as sum of integrals}) into Eq.~(\ref{equation: fat tail integrated}) gives
\begin{equation}
\label{equation: fat tail summed postitve q}
	\tau_0^2 = \frac{2}{\pi R}(1-2R)^2 + 8\lim_{T\to\infty} \frac{1}{T} \sum_{m=1}^{\infty} \int_{\beta_{m+1,m}^{\rm min}(R)}^{\beta_{m+1,m}^{\rm max}(R)} \frac{\text{d}\beta}{2\pi} \int_{b_{m+1,m}^{\rm min}(\beta,R)}^{b_{m+1,m}^{\rm max}(\beta,R)} \frac{\text{d}b}{2R} \tau^{*3}_{m+1,m}(b,\beta,R)\text{H}\left[ T - \tau^*_{m+1,m}(b,\beta,R) \vphantom{\frac{1}{2}} \right] .
\end{equation}
Using geometrical considerations, this time we have $\tau^*_{m+1,m}(b,\beta,R)\simeq[2(m+1/2)^2+(1/\sqrt{2}-2R)^2]^{1/2}\simeq\sqrt{2}m$ for large $m$, so Eq~(\ref{equation: fat tail summed postitve q})'s sum is truncated at (the closest integer to) $T/\sqrt{2}$
\begin{equation}
\label{equation: fat tail summed postitve q truncated}
	\tau_0^2 = \frac{2}{\pi R}(1-2R)^2 + \frac{8}{\sqrt{2}}\lim_{T\to\infty} \frac{\sqrt{2}}{T} \sum_{m=1}^{T/\sqrt{2}} \int_{\beta_{m+1,m}^{\rm min}(R)}^{\beta_{m+1,m}^{\rm max}(R)} \frac{\text{d}\beta}{2\pi} \int_{b_{m+1,m}^{\rm min}(\beta,R)}^{b_{m+1,m}^{\rm max}(\beta,R)} \frac{\text{d}b}{2R} \tau^{*3}_{m+1,m}(b,\beta,R) .
\end{equation}
After loosing the sum as was done for the two open horizons case, Eq.~(\ref{equation: fat tail summed postitve q truncated}) becomes
\begin{equation}
	\tau_0^2 = \frac{2}{\pi R}(1-2R)^2 + \frac{8}{\sqrt{2}}\lim_{T\to\infty} \int_{\beta_{T+1,T}^{\rm min}(R)}^{\beta_{T+1,T}^{\rm max}(R)} \frac{\text{d}\beta}{2\pi} \int_{b_{T+1,T}^{\rm min}(\beta,R)}^{b_{T+1,T}^{\rm max}(\beta,R)} \frac{\text{d}b}{2R} \left[2\left(T+\frac{1}{2}\right)^2+\left(\frac{1}{\sqrt{2}}-2R\right)^2\right]^{3/2} ,
\end{equation}
which then yields Eq.~(\ref{equation: fat tail tau zero extended}). We are left with calculating the remaining parameters for the flag case. For the mean time between collisions we have
\begin{equation}
\label{equation: tau as sum of integrals positive q}
	\langle\tau\rangle = 8\sum_{n=0}^{\infty} \int_{\beta_{n,1}^{\rm min}(R)}^{\beta_{n,1}^{\rm max}(R)} \frac{\text{d}\beta}{2\pi} \int_{b_{n,1}^{\rm min}(\beta,R)}^{b_{n,1}^{\rm max}(\beta,R)} \frac{\text{d}b}{2R}\tau^*_{n,1}(b,\beta,R) + 8\sum_{m=1}^{\infty} \int_{\beta_{m+1,m}^{\rm min}(R)}^{\beta_{m+1,m}^{\rm max}(R)} \frac{\text{d}\beta}{2\pi} \int_{b_{m+1,m}^{\rm min}(\beta,R)}^{b_{m+1,m}^{\rm max}(\beta,R)} \frac{\text{d}b}{2R}\tau^*_{m+1,m}(b,\beta,R) .
\end{equation}
Notice that if one truncates the first sum in Eq.~(\ref{equation: tau as sum of integrals positive q}) at $T$, one then needs to truncate the second sum at $T/\sqrt{2}$. Using $T=500$, we obtain $\langle\tau\rangle\approx1.1947$. For $C_{\psi}$ we write
\begin{equation}
	\ln(T) = \left[1-q\vphantom{\frac{1}{1}}\right]\ln(T) + q\ln\left(\frac{T}{\sqrt{2}}\right) + \frac{1}{2}\ln(2)q \Rightarrow \lim_{T\to\infty} \left\{ \ln(T) - \left[1-q\vphantom{\frac{1}{1}}\right] \sum_{n=1}^T\frac{1}{n} -q\sum_{m=1}^{T/\sqrt{2}}\frac{1}{m} \right\} = \frac{1}{2}\ln(2)q - \gamma ,
\end{equation}
and consequently
\begin{align}
\label{equation: Cpsi as sum of integrals positive q truncated}
	C_{\psi} = \lim_{T\to\infty} \exp & \left\{ \frac{1}{2}\ln(2)q - \frac{3}{2} - \ln\left( \tau_0 \right) - 8\int_{\beta_{0,1}^{\rm min}(R)}^{\beta_{0,1}^{\rm max}(R)} \frac{\text{d}\beta}{2\pi} \int_{b_{0,1}^{\rm min}(\beta,R)}^{b_{0,1}^{\rm max}(\beta,R)} \frac{\text{d}b}{2R} \frac{\tau^{*2}_{0,1}(b,\beta,R)}{\tau_0^2} \vphantom{\sum_{m=1}^{T/\sqrt{2}}} \right. \nonumber \\
	-\sum_{n=1}^T &\left[8 \int_{\beta_{n,1}^{\rm min}(R)}^{\beta_{n,1}^{\rm max}(R)} \frac{\text{d}\beta}{2\pi} \int_{b_{n,1}^{\rm min}(\beta,R)}^{b_{n,1}^{\rm max}(\beta,R)} \frac{\text{d}b}{2R} \frac{\tau^{*2}_{n,1}(b,\beta,R)}{\tau_0^2} - \frac{1-q}{n} \right] \nonumber \\
	-\sum_{m=1}^{T/\sqrt{2}} &\left[8 \int_{\beta_{m+1,m}^{\rm min}(R)}^{\beta_{m+1,m}^{\rm max}(R)} \frac{\text{d}\beta}{2\pi} \int_{b_{m+1,m}^{\rm min}(\beta,R)}^{b_{m+1,m}^{\rm max}(\beta,R)} \frac{\text{d}b}{2R} \frac{\tau^{*2}_{m+1,m}(b,\beta,R)}{\tau_0^2} - \frac{q}{m} \right] \left. \vphantom{\sum_{m=1}^{T/\sqrt{2}}}\right\} .
\end{align}
The behavior displayed in Eq.~(\ref{equation: richardson justification}) was checked to be valid for both sums in Eq.~(\ref{equation: Cpsi as sum of integrals positive q truncated}). Thus, we take $T=40$ and use Richardson extrapolation for each sum separately by extrapolating over the last four partial sums (for the second sum we use terms for which $m \le 28 \approx T/\sqrt{2}$), and obtain $C_{\psi}\approx1.5250\times 10^{-2}$.

\subsection{Stadium channel model}

We use an identical way as for the Lorentz gas to compute $\tau_0$ analytically. Since $\tau_{2n,D}^*(a,\alpha)$ is the traveling distance to the $n$th upper semicircle, it is clear that $\tau_{2n,D}^*(a,\alpha)\simeq\sqrt{(2n+2)^2+D^2}$ when $n$ is large. Plugging Eq.~(\ref{equation: psi as sum of integrals pipe}) into Eq.~(\ref{equation: fat tail integrated}) and employing similar manipulations as before, we have
\begin{equation}
\label{equation: fat tail summed pipe}
	\tau_0^2 = 4\lim_{T\to\infty} \frac{1}{T} \sum_{n=0}^{\infty} \int_{\alpha_{2n,D}^{\rm min}}^{\alpha_{2n,D}^{\rm max}} \frac{\text{d}\alpha}{2\pi} \int_{a_{2n,D}^{\rm min}(\alpha)}^{a_{2n,D}^{\rm max}(\alpha)} \frac{\text{d}a}{2} \tau^{*3}_{2n,D}(a,\alpha) \text{H}\left[ T - \tau^*_{2n,D}(a,\alpha) \vphantom{\frac{1}{2}} \right] ,
\end{equation}
Since $\sqrt{(2n+2)^2+D^2}\simeq2n$ for large $n$, the Heaviside function truncates the sum in Eq.~(\ref{equation: fat tail summed pipe}) at $n=T/2$. Thus
\begin{equation}
	\tau_0^2 = 2\lim_{T\to\infty} \frac{2}{T} \sum_{n=0}^{T/2} \int_{\alpha_{2n,D}^{\rm min}}^{\alpha_{2n,D}^{\rm max}} \frac{\text{d}\alpha}{2\pi} \int_{a_{2n,D}^{\rm min}(\alpha)}^{a_{2n,D}^{\rm max}(\alpha)} \frac{\text{d}a}{2} \tau^{*3}_{2n,D}(a,\alpha) ,
\end{equation}
and consequently
\begin{equation}
	\tau_0^2 = 2\lim_{T\to\infty} \int_{\alpha_{2T,D}^{\rm min}}^{\alpha_{2T,D}^{\rm max}} \frac{\text{d}\alpha}{2\pi} \int_{a_{2T,D}^{\rm min}(\alpha)}^{a_{2T,D}^{\rm max}(\alpha)} \frac{\text{d}a}{2} \left[\sqrt{(2T+2)^2+D^2}\right]^3 .
\end{equation}
Now the integrals can be easily performed. After evaluating the limit we get a closed expression for $\tau_0$, Eq.~(\ref{equation: fat tail tau zero pipe}). For the mean time between collisions, we write
\begin{equation}
	\langle\tau\rangle = 4\int_{\alpha_{0,0}^{\rm min}}^{\alpha_{0,0}^{\rm max}} \frac{\text{d}\alpha}{2\pi} \int_{a_{0,0}^{\rm min}(\alpha)}^{a_{0,0}^{\rm max}(\alpha)} \frac{\text{d}a}{2} \tau^*_{0,0}(a,\alpha) \vphantom{\frac{1}{2}} + 4\sum_{n=0}^{\infty} \int_{\alpha_{2n,D}^{\rm min}}^{\alpha_{2n,D}^{\rm max}} \frac{\text{d}\alpha}{2\pi} \int_{a_{2n,D}^{\rm min}(\alpha)}^{a_{2n,D}^{\rm max}(\alpha)} \frac{\text{d}a}{2} \tau^*_{2n,D}(a,\alpha) ,
\end{equation}
We truncate the sum at $T=500$, and obtain $\langle\tau\rangle\approx 2.57016$. For $C_{\psi}$ we have
\begin{align}
	C_{\psi} = \lim_{T\to\infty} \exp\left\{ \vphantom{\sum_{n=0}^{T/2}} \right. \ln\left( \frac{2}{\tau_0} \right) &- \frac{3}{2} - 4\int_{\alpha_{0,0}^{\rm min}}^{\alpha_{0,0}^{\rm max}} \frac{\text{d}\alpha}{2\pi} \int_{a_{0,0}^{\rm min}(\alpha)}^{a_{0,0}^{\rm max}(\alpha)} \frac{\text{d}a}{2} \frac{\tau^{*2}_{0,0}(a,\alpha)}{\tau_0^2} \vphantom{\frac{1}{2}} - 4\int_{\alpha_{0,D}^{\rm min}}^{\alpha_{0,D}^{\rm max}} \frac{\text{d}\alpha}{2\pi} \int_{a_{0,D}^{\rm min}(\alpha)}^{a_{0,D}^{\rm max}(\alpha)} \frac{\text{d}a}{2} \frac{\tau^{*2}_{0,D}(a,\alpha)}{\tau_0^2} \nonumber \\
	&-\left. \sum_{n=1}^{T/2} \left[4 \int_{\alpha_{2n,D}^{\rm min}}^{\alpha_{2n,D}^{\rm max}} \frac{\text{d}\alpha}{2\pi} \int_{a_{2n,D}^{\rm min}(\alpha)}^{a_{2n,D}^{\rm max}(\alpha)} \frac{\text{d}a}{2} \frac{\tau^{*2}_{2n,D}(a,\alpha)}{\tau_0^2} - \frac{1}{n} \right] \right\} .
\end{align}
Substituting $\tau_{2n,D}^*(a,\alpha)\simeq \sqrt{(2n+2)^2+D^2}$ for $n\gg 1$ and evaluating the integrals, we observe the following behavior
\begin{equation}
	s_n \simeq 4 \int_{\alpha_{2n,D}^{\rm min}}^{\alpha_{2n,D}^{\rm max}} \frac{\text{d}\alpha}{2\pi} \int_{a_{2n,D}^{\rm min}(\alpha)}^{a_{2n,D}^{\rm max}(\alpha)} \frac{\text{d}a}{2} \frac{(2n+2)^2+D^2}{\tau_0^2} - \frac{1}{n} \simeq \frac{A_1}{n^2} + \frac{A_2}{n^3} + \frac{A_3}{n^4} ,
\end{equation}
where $A_1,A_2,A_3$ are some $D$-dependent coefficients. This behavior suggests that the partial sum $S_T$ goes like
\begin{align}
	S_T = \sum_{n=1}^T s_n \simeq l + \frac{\tilde{A}_1}{T} + \frac{\tilde{A}_2}{T^2} + \frac{\tilde{A}_3}{T^3} ,
\end{align}
for large $T$, where $l$ is the desired limit. Once again, we employ the Richardson extrapolation method. For $T=60$, we get $C_{\psi}\approx 1.0903\times 10^{-5}$ by summing terms up to $n=T/2$ and extrapolating over the $S_{27},S_{28},S_{29},S_{30}$ partial sums.

\end{widetext}

\end{document}